\documentclass[aps,superscriptaddress,showpacs,prd,nofootinbib,eqsecnum]{revtex4-1}

\usepackage{graphics} 
  
\usepackage{xcolor}
\usepackage{epsfig} 
\input{epsf}
\usepackage{natbib} 

\usepackage[colorlinks=true,citecolor=blue,linkcolor=blue,urlcolor=blue]{hyperref}

\setcitestyle{square, numbers, comma, sort&compress}

\def\be{\begin{equation}}
\def\ee{\end{equation}}
\def\bea{\begin{eqnarray}}
\def\eea{\end{eqnarray}}
\def\ms{\overline {\rm MS}}

\begin{document} 
 
\title{Renormalization group improved pressure for hot and dense quark matter }

\author{Jean-Lo\"{\i}c Kneur}\email{jean-loic.kneur@umontpellier.fr}
\affiliation{Laboratoire Charles Coulomb (L2C), UMR 5221 CNRS-Universit\'{e} Montpellier, 34095 Montpellier, France}  
\author{Marcus Benghi Pinto} \email{marcus.benghi@ufsc.br}
\affiliation{Departamento de F\'{\i}sica, Universidade Federal de Santa
  Catarina, 88040-900 Florian\'{o}polis, SC, Brazil}
 \author{Tulio E. Restrepo} \email{tulio.restrepo@posgrad.ufsc.br}
\affiliation{Departamento de F\'{\i}sica, Universidade Federal de Santa
  Catarina, 88040-900 Florian\'{o}polis, SC, Brazil}
%\affiliation{CFisUC - Center for Physics of the University of Coimbra, Department of Physics, Faculty of Sciences and Technology, University of Coimbra, 3004-516 Coimbra, Portugal}  
 
 \begin{abstract} 
 We apply the renormalization group optimized perturbation theory (RGOPT) to evaluate
 the quark contribution to the QCD pressure at finite temperatures and baryonic 
 densities, at next-to-leading order (NLO).
  Our results are compared to NLO and state-of-the-art higher orders of 
 standard perturbative QCD (pQCD) and hard thermal loop perturbation theory 
 (HTLpt). The RGOPT provides an  all order resummed pressure in a well-defined  
 approximation, exhibiting a drastically better remnant renormalization
 scale dependence than pQCD, thanks to built-in renormalization group invariance consistency.  
  At NLO, upon simply adding to the RGOPT-resummed quark contributions the purely perturbative
 NLO glue contribution, our results show a remarkable agreement with ab initio lattice simulation 
 data for temperatures $0.25 \lesssim T \lesssim 1 \, {\rm GeV}$, 
 with a remnant scale dependence drastically reduced as compared to HTLpt.
 \end{abstract} 
 
 
\maketitle
 
\section{Introduction} 
The complete prediction of the phase diagram describing strongly interacting matter transitions  represents one of 
the major theoretical challenges in contemporary particle physics, despite the  enormous progress achieved  by 
lattice QCD (LQCD) numerical simulations. The main reason is that the well documented sign problem \cite{sign}, 
which arises when finite chemical
potential ($\mu$) values are considered, prevents LQCD to be reliably applied at intermediate to high finite 
baryonic densities, while at low densities the problem may be circumvented, e.g., by performing a Taylor expansion 
around vanishing chemical potential results. In particular, within the latter regime  LQCD has been very successful
in predicting \cite{lattice} that a crossover occurs at a pseudocritical temperature close to $T_{pc} \approx 155\, {\rm MeV}$ 
when $\mu=0$.  One alternative to describe the low temperature-high density domain is to employ effective 
quark theories \cite{CET}, or the Nambu--Jona-Lasinio (NJL) model \cite{njl}, evaluating physical quantities within 
some analytical more nonperturbative framework (e.g., mean field theory, MFT). This approach predicts that the (chiral) phase transition 
at low-$T$ and finite $\mu$ is of the first kind \cite {buballa} so that, as a byproduct, one should observe 
a critical end point (CP)
signalled by  a second order phase transition taking place at intermediate values of $T$ and $\mu$ where the first order
transition boundary terminates. This intriguing possibility is about to be tested in heavy-ion collisions experiments 
by decreasing the beam energy, $\sqrt {s_{NN}}$, so that the baryonic density increases. In view of these experiments it is 
an unfortunate situation that theoretical predictions using the full QCD machinery cannot be consistently carried out 
with the currently available nonperturbative techniques. 

As already emphasized LQCD is plagued by the sign problem while analytical tools such 
as the large-$N$ approximation (which is related to MFT) may produce misleading results at criticality. 
More analytical alternatives to LQCD can partly address the
deconfinement and/or chiral symmetry restoration at finite temperature and/or density: 
typically, some extensions of the NJL model\cite{buballa,NJLCosta} or Polyakov NJL (PNJL)\cite{PNJL,PNJLCosta}, 
or other approaches incorporating more basic QCD dynamics in well-defined approximations, 
like the Dyson-Schwinger equations (see e.g. \cite{DSE,DSErecent}), the functional Renormalization Group (see e.g. \cite{fRG}), 
or other approaches\cite{others}.
On another side, standard thermal perturbation theory (PT) 
is unreliable at the relevant temperature and chemical potential ranges. Indeed,
despite the asymptotic 
freedom (AF) property, its convergence can only be achieved at temperatures many orders of magnitude larger than the 
critical one. Even at intermediate temperatures,
it is well-known that thermal PT is plagued by severe infrared divergences from bosonic zero modes, and has to be resummed 
to be more compatible with strong coupling regimes (for pedagogical reviews and lecture notes
see, e.g., Refs. \cite{Trev,laine} and the very recent Ref. \cite{HTLrev2020}).
Yet, even the state-of-the-art, highest available order thermal PT \cite{pQCD4L,pQCDmu4L}, 
that incorporates a suitable resummation of infrared divergences, becomes more poorly 
accurate at moderate to low temperatures.
A very successful alternative resummation
method is to systematically expand from the start 
about a quasiparticle mass \cite{spt,HTLbasic,HTLpt}, that also more directly avoids 
infrared divergences apart from improving convergence issues.
(One should keep in mind however that not all infrared divergences can be cured by such resummations: 
as is well-known the screening of static magnetic 
fields\cite{Linde} is an additional nonperturbative phenomenon occuring in Yang-Mills theory, intrinsically limiting
in practice the weak coupling expansion to maximal order $\alpha_S^3\ln \alpha_S$).
The expansion about a quasiparticle mass is actually close to analytical resummation approaches also 
used at zero-temperature, reminiscent of the traditional Hartree approximation and its 
variational generalizations, suitable to tackle the infrared divergence issues of massless theories.
Basically one essentially deforms the 
original Lagrangian by a Gaussian mass term to be treated as an interaction, 
defining a modified perturbative expansion 
leading to a sequence of (variationally improved) approximations at successive orders.

The latter approaches appear under various names in the literature, such 
as optimized perturbation theory (OPT)~\cite{pms,opt_phi4,opt_qcd} (as we dub it here),
linear $\delta$ expansion (LDE)~\cite{lde},
variational perturbation theory (VPT)~\cite {vpt}, or screened perturbation theory (SPT)~\cite{spt,spt4L}
in the thermal context. Remark 
that adding a Gaussian term does not change the polynomial structure of the theory so that  
the process is compatible with the usual renormalization procedure.
Already at NLO one usually goes beyond  the simple Hartree 
approximation since the variational mass is dressed by incorporating different resummed topologies 
(exchange terms, vertex corrections, etc) order by order. 
Moreover, at leading order the OPT has the welcome property of exactly reproducing 
large-$N$ results~\cite{npb}.
As discussed, e.g., in Ref.~\cite{prdGN} this technique has been used to describe successfully a variety of
physical situations, involving phase transitions in different models. 
On the other hand, for thermal theories,  the SPT method has been generalized
over the past two decades in order to be compatible with Yang-Mills theories. This generalization was made 
possible thanks to the hard thermal loop (HTL) gauge-invariant effective Lagrangian 
originally built by Braaten and Pisarski~\cite{HTLbasic}, 
 consistently embedding HTL
contributions, Landau damping and a screening gluon thermal mass term, with momentum-dependent self-energies and HTL-dressed vertices. 
The high temperature expansion based on the HTL Lagrangian, known as hard thermal loop perturbation 
theory (HTLpt) \cite{HTLpt}, has been employed in a series of applications up to NNLO (three-loops), 
to describe the QCD thermodynamics, considering both the glue \cite{HTLptg3L} and 
quark \cite{HTLptqcd2L,HTLptDense3L,HTLptqcd3L} sectors at 
finite temperatures and baryonic densities. Given the intrinsic 
technical difficulties associated with the HTLpt evaluations, the NNLO state-of-the-art calculations
performed typically in Refs. \cite{HTLptDense3L,HTLptqcd3L} represents a remarkable achievement.
 Unfortunately it is worth noting a serious remnant issue, also plaguing standard PT 
but not sensibly reduced in HTLpt: 
namely the sensitivity to the arbitrary renormalization scale is observed to substantially {\em increase} when 
higher orders are considered. 
More precisely, as compared to PT the NNLO HTLpt predictions in Refs.\cite{HTLptDense3L,HTLptqcd3L} are very close 
to the lattice results for temperatures
down to $T \gtrsim 2\, T_{pc}$ for  the commonly chosen ``central'' renormalization scale choice,
$M = 2\pi \sqrt  {T^2 + \mu^2/\pi^2}$. However, even a moderate scale variation of a factor 2 dramatically affects the pressure
and related thermodynamical quantities by relative variations of order 1 or more. 
It has been argued \cite{HTLptqcd3L} that resumming logarithms may help to improve the situation but, as explained in Refs.\cite{prlphi4,prdphi4}, 
it appears that the lack of renormalization group (RG) invariance is more basically rooted within the HTLpt approach.\\
More recently an alternative combining the OPT framework with RG properties has been proposed: 
the renormalization group optimized perturbation theory (RGOPT)\cite{JLGN,JLalphas,prlphi4,prdphi4}. 
The main novelty is that it restores RG invariance
at all stages of the calculation, in particular when fixing
the arbitrary variational mass parameter.
At vanishing temperatures it has been used in QCD up to high (three and four-loop) orders
to estimate the basic coupling $\alpha_s$~\cite{JLalphas}, 
predicting values very compatible with the world averages \cite{PDG2018}. 
Also accurate values of the (vacuum) quark condensate were obtained 
at four-loop\cite{JLcond} and five-loop\cite{JLcond2} orders. 
Concerning thermal theories the RGOPT has been applied 
to the simpler scalar $\phi^4$ model \cite{prlphi4,prdphi4} at NLO, 
as well as to the nonlinear sigma model (NLSM) \cite{nlsm}.
In these thermal applications the RGOPT and PT/SPT predictions for the pressure have 
been compared, showing how the former approximation  ameliorates  the generic
residual scale dependence of thermal perturbation theories at increasing perturbative orders. 
More recently we have evaluated the quark contribution to the QCD pressure at two-loop (NLO) at finite densities 
and vanishing temperatures, showing how the method improves over perturbative QCD (pQCD)~\cite{prdCOLD}. 
In the present work we extend our approach to include the effects of a thermal bath. Note that
applying the RGOPT readily to the glue contributions 
is beyond the present scope, due to some specific technical difficulties as briefly explained below
(work in this direction is in progress \cite{gluons}). 
Therefore in the present application the RGOPT resummation will 
be applied strictly only to the quark sector, while the gluons 
will be treated as in standard (NLO) PT. In the end both contributions will be combined  
 in order to produce our complete final prediction for the NLO QCD pressure.  

The paper
is organized as follows.  In the next section we briefly review our starting point, the perturbative expressions
considered for the (massive) quark pressure at NLO, for which the basic RGOPT construction is recalled. 
In Sec. III the RGOPT is precisely defined for the 
quark pressure up to NLO (two-loop). Details of our two possible prescriptions at NLO 
are described in Sec. IV (that may be skipped by the reader only interested in 
the main results). Then Sec.V illustrates our main results for the pressure, 
both for the pure quark sector and for the full QCD one.
We compare our results with the NLO and state-of-the-art higher orders of both PT and HTLpt, and also to lattice data
for the complete QCD pressure. Sec. VI contains our conclusions and perspectives. Finally, three appendices 
specify some formulas and additional details used in our analysis.
\section {Quark Contribution to the QCD  Pressure}
\subsection {RG invariant perturbative pressure}
\begin{figure}[htb]
 \includegraphics{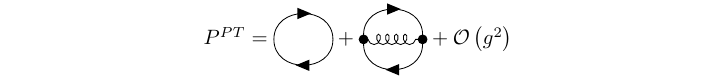}
\caption{Feynman diagrams contributing to the perturbative quark pressure up to NLO ${\cal O}(g)$. }
\label{Fig0}
\end{figure}
At two-loop order-$g$~\footnote{In all what follows we normalize 
for convenience the (running) coupling in the $\ms$-scheme as $g(M)\equiv 4\pi\alpha_S(M)$.}, 
the contribution of (massive) quarks to the QCD perturbative pressure (the Feynman diagrams displayed 
in Fig. \ref{Fig0}) can be obtained by combining the vacuum ($T=\mu=0$) 
results of Ref.\cite {JLcond} and the $T,\mu \ne 0$ results of Refs.\cite{kapusta-gale,laine2}. 
Considering the case of degenerate masses $m_u=m_d=m_s\equiv m$, the renormalized pressure 
in the $\ms$ renormalization scheme, normalized per flavor, is
\begin{eqnarray}
\frac{P^{PT}_1}{N_f \,N_c}&=&-\frac{m^4}{8 \pi^2} \left(\frac{3}{4}-L_m\right)+2T^4 J_1\left(\frac{m}{T},\frac{\mu}{T}\right)
-3g \frac{m^4}{2\left(2\pi\right)^4} C_F \left(L_m^2-\frac{4}{3}L_m+\frac{3}{4}\right)\nonumber \\
&-&g C_F \left \{ \left[\frac{m^2}{4\pi^2}  \left(2-3L_m \right) + \frac{T^2}{6}  
\right] T^2 J_2\left(\frac{m}{T},\frac{\mu}{T}\right) +\frac{T^4}{2} J^2_2\left(\frac{m}{T},\frac{\mu}{T}\right) +m^2 
T^2 J_3 \left(\frac{m}{T},\frac{\mu}{T}\right) \right \} \;,
\label{PPT}
\end{eqnarray}
where $g\equiv g(M)$, $L_m\equiv \ln(m/M)$, $M$ is the arbitrary renormalization scale, 
$C_F = (N_c^2-1)/(2 N_c)$, $N_c=3$, and $N_f=3$.
The in-medium and thermal effects are included in the (dimensionless) single integrals:
\begin{equation}
 J_1\left(\frac{m}{T},\frac{\mu}{T}\right)=\int\frac{ d^3{\bf \hat p}}{\left(2\pi\right)^3}
 \left\lbrace\ln\left[1+e^{-\left(E_p+\frac{\mu}{T}\right)}\right]+\ln\left[1+e^{-\left(E_p-\frac{\mu}{T}\right)}\right]\right\rbrace \;,
\label{J1def}
 \end{equation}
with ${\bf \hat p}\equiv {\bf p}/T$, $E_p=\sqrt{{\bf \hat p}^2 +\frac{m^2}{T^2}}$,  
\begin{equation}
 J_2\left(\frac{m}{T},\frac{\mu}{T}\right)=\int\frac {d^3 {\bf \hat p}}{\left(2\pi\right)^3}\frac{1}{E_p}
 \left[f^{+}(E_p)+f^-(E_p)\right]\;,
 \label{J2def}
\end{equation}
as well as in the double integral (after performing exactly the angular integration over $p\cdot q/(|p||q|)$)
\begin{equation}
J_3\left(\frac{m}{T},\frac{\mu}{T}\right)=\frac{1}{(2\pi)^4} \int^\infty_0 \int^\infty_0 \frac{d\hat p\, \hat p\, d\hat q 
\,\hat q}{E_p E_q} 
\left\{  \Sigma_+ \ln\left[\frac{E_p E_q -\frac{m^2}{T^2}-\hat p \hat q}{E_p E_q -\frac{m^2}{T^2}+\hat p \hat q}\right] +
\Sigma_- \ln\left[\frac{E_p E_q +\frac{m^2}{T^2}+\hat p \hat q}{E_p E_q +\frac{m^2}{T^2} -\hat p \hat q}\right]
\right \} \;,
\label{J3def}
\end{equation}
where
\begin{eqnarray}
 \Sigma_{\pm}&=&f^+\left(E_p\right)f^{\pm}\left(E_q\right)+f^-\left(E_p\right)f^\mp\left(E_q\right),\\
\end{eqnarray}
in terms of the Fermi-Dirac distributions  for anti-quarks ($+$ sign) and quarks ($-$ sign), 
\begin{equation}
f^\pm(E_p)=\frac{1}{1+e^{(E_p\pm \frac{\mu}{T})}} \;,
\end{equation}
where $\mu$ represents the quark chemical potential, which relates to the baryonic chemical potential via $\mu_B = 3 \mu$. 

In the present work we consider the case of symmetric quark matter and so do not distinguish the chemical potentials 
associated with the different flavors ($\mu_s=\mu_u=\mu_d\equiv \mu$). The generalization to the case of 
chemical equilibrium needed to impose, e.g., $\beta$ equilibrium should be straightforward.
Also relevant for our purpose and comparisons is the well-known resulting two-loop pressure expression for 
strictly massless quarks (that simplifies
considerably since the $J_i$ integrals reduce to simple analytical expressions in this case, given 
for completeness in Appendix \ref{ApphT}):
\begin{equation}
\frac{P^{PT}_1(m\to 0)}{P_{SB}(T,\mu)} = 1 -\frac{25 g(M)}{42\pi^2}\, 
\left ( \frac{1 + \frac{72}{5}  {\hat \mu}^2 + \frac{144}{5}  {\hat \mu}^4}
{1+  \frac{120}{7} {\hat \mu}^2 + \frac{240}{7} {\hat \mu}^4} \right )
\label{PPTm0}
\end{equation}
with $ {\hat \mu} = \mu / (2\pi T)$. 
The Stefan-Boltzmann (SB) ideal gas limit reads 
\begin{equation}
P_{SB}(T,\mu)= T^4 N_f N_c \left ( \frac {7 \pi^2}{180} \right )
\left ( 1 + \frac{120}{7} {\hat \mu}^2 + \frac{240}{7} {\hat \mu}^4 \right ) \;.
\label{Pqsb}
\end{equation}
Coming back to the massive quark case, we next define the standard homogenous RG operator, 
\begin{equation}
M \frac{d}{d M}= M \,\partial_M + \beta(g)\,\partial_g  - m\,\gamma_m(g) \, \partial_m \;,
\label{RG}
\end{equation}
where our normalization convention for the QCD $\beta$-function and anomalous mass dimension $\gamma_m$ is
\begin{equation}
 \beta\left(g\right)=-2b_0g^2-2b_1g^3+\mathcal{O}\left(g^4\right) \;,
 \end{equation}
 \begin{equation}
 \gamma_m\left(g\right)=\gamma_0 g+\gamma_1g^2+\mathcal{O}\left(g^3\right)\;,
\end{equation}
where to two-loop order,
 \begin{equation}
  \left(4\pi\right)^2 b_0= 11-\frac{2}{3}N_f ,
 \label{b0def}
 \end{equation}
\begin{equation}
 \left(4\pi\right)^4 b_1 = 102-\frac{38}{3}N_f ,
\label{b1def}
\end{equation}
\begin{equation}
 \gamma_0 =\frac{1}{2\pi^2} \;,\;\;
\left(4\pi\right)^4 \gamma_1= \frac{404}{3}-\frac{40}{9}N_f \;.
\label{gamdef}
\end{equation}
As is well known, the {\em massive} pressure 
(equivalently the massive free energy) is lacking perturbative RG-invariance: it can be easily checked that
applying Eq.(\ref{RG}) to  Eq.(\ref{PPT})
leaves a remnant term starting at {\em leading} order. 
Now to turn Eq.(\ref{PPT}) into a (perturbatively) RG invariant quantity, 
we proceed as in Refs.\cite{JLcond,prlphi4,prdphi4} (or closer to the present case, as in Ref.\cite{prdCOLD}), 
by subtracting a finite zero-point contribution,
\begin{equation}
\frac{P^{RGPT}}{N_c N_f} = \frac{P^{PT}}{N_c N_f} - \frac{m^4}{g} \sum_k s_k g^k \; ,
\label{PRGPT}
\end{equation}
where the $s_i$ are determined at successive perturbative orders so that 
\begin{equation}
M \frac{d}{d M} \left (\frac{P^{RGPT}}{N_c N_f} \right )  ={\cal O}(g^2 m^4) \,,
\label{RGeq}
\end{equation}
up to neglected higher order terms. 
Since our evaluations are being carried up to NLO, to restore perturbative RG invariance
at this order it is sufficient to add the first two 
$s_0$, $s_1$ coefficients that involve the coefficients of $\beta(g)$, $\gamma_m(g)$ through Eq.(\ref{RG}). 
One finds explicitly \cite{JLcond,prdCOLD}
\begin{equation}
 s_0=-\left[(4\pi)^2 (b_0-2\gamma_0)\right]^{-1} ,
 \label{s0def}
 \end{equation}
 \begin{equation}
 s_1=-\frac{1}{4}\left[\frac{b_1-2\gamma_1}{4(b_0-2\gamma_0)} -\frac{1}{12\pi^2}\right ].
 \label{s1def}
\end{equation}
\subsection {Implementing the RGOPT for the quark pressure} \label{secRGOPT}
 The RGOPT construction\cite{JLalphas,prlphi4,prdCOLD}  may be summarized as the following successive steps:
\begin{itemize}
\item 1. First one restores perturbative RG invariance of the massive 
theory, following the procedure above described, giving at NLO $P^{RGPT}(m,g)$ in Eqs.(\ref{PRGPT}), (\ref{s0def}),(\ref{s1def}). 
The subtraction contributions can be viewed as 
extending the Lagrangian by vacuum terms $\sim (m^4/g) H(g)$, that could be equivalently introduced at the bare 
level\cite{prdphi4,JLcond2}, although the above prescription working directly with renormalized quantities is most convenient.
At this stage, upon using commonly defined NLO running coupling and masses from Eqs.(\ref{b0def})--(\ref{gamdef}), it could easily 
be checked that Eq.(\ref{PRGPT}) has a remnant scale dependence only of higher order $\sim m^4 g^2$ by construction. 
\item 2. Next the RGOPT requires to variationally deform the Lagrangian, by rescaling the coupling 
(consistently for every standard interaction terms), and modifying quark mass terms, as 
\begin{equation}
{\cal L}_{QCD}^{RGOPT}= {\cal L}_{QCD}(g\to \delta g,  m\to m (1-\delta)^a) ,
 \label{Lint}
\end{equation}
%where $ {\cal L}_{QCD}(g,m_q)$ is the renormalized QCD Lagrangian 
$m$ being now an arbitrary mass. 
 Eq.(\ref{Lint}) is to be viewed as a convenient bookkeeping 
prescription actually performed on an already renormalized quantity $P(m,g)$, thus obtained from standard perturbative expansion and 
appropriate renormalizations of the {\em massive} theory. 
Concerning the pressure it amounts to do those substitutions within $P^{RGPT}(m,g)$
of Eq.(\ref{PRGPT}), thus incorporating the previously obtained vacuum subtractions~\footnote{ For 
overall RG consistency it is crucial to perform first the perturbative RG-restoring procedure as in Eq.(\ref{PRGPT}), before 
modifying the perturbative expansion according to Eq.(\ref{Lint}), as the two procedures do not commute.}.
\item 3. The resulting expression from Eq.(\ref{Lint}) is expanded in powers of $\delta$, at the same perturbative order considered, 
and next setting $\delta\to 1$ to recover the massless limit. At any finite order this leaves a residual $m$-dependence, 
that can be fixed by a stationarity criterion\cite{pms}, the mass optimization prescription (MOP):
 \begin{equation}
\frac{\partial}{\partial m} P^{RGOPT}({\cal O}(\delta^k),\delta \to 1) \Bigr |_{\overline m} \equiv 0 \;.
\label{mop}
\end{equation}
Eq.(\ref{mop}) is however not the sole nor the most compelling prescription once 
RG invariance properties are considered. 
Moreover, while in related OPT/SPT or HTLpt approaches the simple (linear) value $a=1$ was used 
in Eq.~(\ref{Lint}) mainly for simplicity, it was observed\cite{JLalphas,prlphi4} 
that step 3 generally spoils RG invariance at leading order, even when Eq.(\ref{Lint}) is performed on a
perturbatively RG invariant pressure like Eq.(\ref{PRGPT}). 
\item 4. Thus in conjunction with step 3, to preserve good RG properties of the variationally modified pressure 
the further RGOPT ingredient is to fix\cite{JLalphas,prlphi4} the exponent $a$ purposely introduced in Eq.(\ref{Lint}): 
we require the $\delta\to 1$ pressure to satisfy, since Eq.(\ref{mop}) is used, the {\it reduced} (massless) RG equation, 
\begin{equation}
 \left [ M \partial_M +\beta(g)\partial_g \right ]P^{RGOPT}_{\delta\to 1} = 0\; .
 \label{RGred}
\end{equation}
For this
it is sufficient to expand to leading order ($\delta^0$),
yielding 
\begin{equation}
 a= \frac{\gamma_0}{2 b_0} \,,
 \label{aQCD}
\end{equation}
that only depends on the universal (scheme-independent) LO RG coefficients\footnote{ At NLO and higher orders one could possibly 
further generalize the interpolation $(1-\delta)^a $ in Eq.(\ref{Lint}) with $\delta^2$ and higher order terms without spoiling 
the crucial LO RG properties entirely fixed by Eq.(\ref{aQCD}). But this would involve extra arbitrary variational parameters 
with no compelling reasons, with no way of fixing those 
so that the remnant NLO scale dependence of ${\cal O}(g^2 m^4)$ would exactly disappear.
Keeping a minimal form with only $a$ in Eq.(\ref{aQCD}) gives the same prescription at successive orders, 
thus more sensible for comparisons.},
in agreement with previous RGOPT applications to which we refer for detailed demonstration \cite{JLalphas,prdphi4,prdCOLD}.
At NLO Eq.(\ref{RGred}) is no longer exactly satisfied by Eq.(\ref{aQCD}) alone, thus it 
can provide an alternative determination of $\overline m$ besides Eq.(\ref{mop}), as will be illustrated 
in concrete NLO applications below. 
Importantly, keeping Eq.(\ref{aQCD}) at higher orders further guarantees 
that the only acceptable solutions are those matching\cite{JLalphas} the perturbative asymptotic
freedom (AF) behavior for $g\to 0$ at $T=0$. 
This simple but compelling criterion often selects a unique solution, even at five-loop
order so far explored~\cite{JLcond2}, in contrast with the related OPT/SPT approaches
where using solely Eq.(\ref{mop}) generates an increasing number of possible solutions at increasing orders.
\end{itemize}
We recall that 
%at this stage that the RGOPT construction steps 1-4 above is rather generic and may be applied, 
%with minor adaptations, both to bosonic or fermionic contributions if relevant
%(provided that their appropriate RG anomalous dimensions are used, if known, for each sectors). 
%applying a similar procedure for the $\phi^4$ model gets rid\cite{prdphi4} of infrared divergences originating at finite temperature 
%from bosonic zero modes, so we anticipate similar properties would hold for the QCD gluon sector, let aside the above mentioned present obstacles to 
%readily apply our construction in this sector. Now 
the quark sector at finite temperature has no zero modes, 
thus strictly there is no need to resum infrared divergences, and in standard thermal PT this sector is commonly treated purely 
perturbatively\cite{pQCD4L,pQCDmu4L}.
Accordingly since our present construction is being performed solely for the quark sector, it is not dealing with thermal infrared divergences 
(that anyhow occur only at NNLO from the gluon sector). 
It nevertheless resums well-defined RG-induced higher order contributions, 
leading to rather drastic differences with NLO pQCD, as will be illustrated below.\\
At lowest nontrivial order $\delta^0$ the resulting RGOPT pressure is given, keeping 
all terms of formally one-loop order, by
\begin{equation}
  \frac{P^{RGOPT}_0}{N_fN_c}=-\frac{2m^4}{\left(4\pi\right)^2} \left(\frac{3}{4}-L_m\right)-m^4\,s_1 +2T^4 
  J_1\left(\frac{m}{T},\frac{\mu}{T}\right)+\frac{m^4}{\left(4\pi\right)^2 g\,b_0} \,.
  \label{P1LRGOPT}
\end{equation}
Remark that the LO $s_0$ coefficient, Eq.(\ref{s0def}), has produced
the last term $\propto 1/b_0$ in Eq.(\ref{P1LRGOPT}) after algebraic simplifications. 
 There is a subtlety here: as Eq.(\ref{s1def}) shows,  $s_1$ involves {\em two-loop} RG coefficients and 
 thus it is not mandatory to restore (perturbative) RG invariance 
 at LO, 
that requires only $s_0\ne 0 $ as  explained. 
 Yet, since $s_1$ enters the pressure  
formally at ${\cal O}(1)$, it appears sensible to include it also
within our one-loop RGOPT result Eq.(\ref{P1LRGOPT}), incorporating in this way
higher order RG properties. 
 (Actually the difference between the LO  
prescriptions with $s_1\ne 0$ or 
taking more simply $s_1=0$ is not drastic). 
At the one-loop level the coupling runs according to the well-known expression
\begin{equation}
g\left(M\right)= \frac{1}{2b_0 \, \ln (M/\Lambda_{\overline{\rm MS}})} .
 \label{g1L}
\end{equation}
 Proceeding similarly at the next RGOPT order, 
 the NLO pressure reads,  after performing steps 1-4 above,
\begin{eqnarray}
\frac{P^{RGOPT}_1}{N_fN_c}&=&-\frac{m^4}{8 \pi^2} \left(\frac{3}{4}-L_m\right)+2T^4\,J_1\left(\frac{m}{T},\frac{\mu}{T}\right)
+\frac{m^4}{\left(2\pi\right)^2}\left(\frac{\gamma_0}{b_0}\right)\left(\frac{1}{2}-L_m\right) \nonumber \\
&+& m^2 \left(\frac{\gamma_0}{b_0}\right) T^2\,J_2\left(\frac{m}{T},\frac{\mu}{T}\right) 
+\frac{m^4}{\left(4\pi\right)^2 b_0}\left\lbrace \frac{1}{g}\left (1-\frac{\gamma_0}{b_0} \right )
+ \left[\left(b_1-2\gamma_1\right)\pi^2 -\frac{\left(b_0-2\gamma_0\right)}{3}\right] \right \} \nonumber \\
 &-&3g  C_F \frac{m^4}{2\left(2\pi\right)^4} \left(L_m^2-\frac{4}{3}L_m+\frac{3}{4}\right)  \nonumber \\
 &-&  g C_F \left \{ \left[\frac{m^2}{4\pi^2}  \left(2-3L_m \right) + \frac{T^2}{6}  
\right]  T^2J_2\left(\frac{m}{T},\frac{\mu}{T}\right) +\frac{T^4}{2} J^2_2\left(\frac{m}{T},\frac{\mu}{T}\right)
+m^2 T^2\,J_3 \left(\frac{m}{T},\frac{\mu}{T}\right) \right \} .
\label{P2LRGOPT}
\end{eqnarray}
The exact two-loop (2L) running coupling, analogue of the one-loop Eq.(\ref{g1L}), is obtained by solving for $g(M)$
the implicit relation (see, e.g., Ref. \cite{jlqcd1})
\begin{equation}
\ln \frac{M}{ \Lambda_{\overline{\text{MS}}} } = \frac{1}{2b_0\, g} +
\frac{b_1}{2b_0^2} \ln \left ( \frac{b_0 g} {1+\frac{b_1}{b_0} g} \right) ,
\label{g2L}
\end{equation}
for a given $\Lambda_{\overline{\text{MS}}}$ value (this also defines the
$\Lambda_{\overline{\text{MS}}}$ basic scale in our normalization conventions). 
In the numerical illustrations below, we will use a value very close to the latest world average value\cite{PDG2018},  
$\Lambda_{\overline{\rm MS}}=  335 \,{\rm MeV}$ for $N_f=3$, that equivalently 
corresponds to $\alpha_s(N_f=3,1.5 \, {\rm GeV})\simeq 0.326$. (NB the latter $\alpha_s$ value precisely 
compares with the one taken in the literature for the NLO PT and HTLpt pressures\cite{HTLptqcd2L}).
\section {RG-optimized resummation}
\subsection{One-loop RGOPT}\label{sec1L} 
 Before proceeding to our most relevant NLO results, derived basically from Eq.(\ref{P2LRGOPT}), 
it is useful to examine the probably more transparent RGOPT features at the lowest nontrivial ($\delta^0$) LO.
We recall that at this order the pressure already satisfies 
the massless RG Eq.(\ref{RGred}) exactly, via the RG-driven exponent Eq.(\ref{aQCD}) of the variationally modified Lagrangian, 
 Eq.(\ref{Lint}).
 Consequently the arbitrary mass $m$ may be fixed only by using the MOP Eq.(\ref{mop}).
The latter acting on the LO pressure Eq.(\ref{P1LRGOPT}) can easily be cast into the form
\begin{equation}
\frac{1}{b_0\,g} +\ln \frac{m^2}{M^2} -1  -16\pi^2 s_1   -
8\pi^2 \frac{T^2}{m^2} J_2\left ({\frac{ m}{T}},{\frac{\mu}{T}} \right ) =0, 
 \label{mopd0}
\end{equation}
whose nontrivial solution gives an RG invariant dressed mass $\overline m(g,T,\mu)$, 
since the combination $ 1/(b_0\,g(M)) +\ln m^2/M^2$ is trivially   
$M$-independent according to Eq.(\ref{g1L}). (NB for more generality we keep $s_1$ unspecified at this stage,
while for numerics below we will take $s_1 \ne 0$ as given by Eq.(\ref{s1def})). 
Once inserting  $\overline m$ in Eq.(\ref{P1LRGOPT}) it produces a (one-loop) exactly RG invariant 
pressure, that takes the compact form:
\begin{equation}
\frac{P^{RGOPT}_0}{N_f N_c} =
2 T^4\, J_1\left (\frac{\overline m}{T},\frac{\mu}{T} \right ) +\frac{T^2}{2}  {\overline m}^2 J_2\left (\frac{\overline m}{T},
\frac{\mu}{T} \right ) 
-\frac{\overline m^4}{32\pi^2}\;,
 \label{Prgoptd0simp}
\end{equation}
where it is understood that $\overline m$ is the nontrivial solution of 
Eq.(\ref{mopd0})~\footnote{Notice that the explicit dependence upon 
$s_1$ cancelled in $P^{RGOPT}_0$ Eq.(\ref{Prgoptd0simp}) upon using 
Eq.(\ref{mopd0}), but the solution $\overline m$ of Eq.(\ref{mopd0}) does depend on $s_1$ as specified.}.
Some properties of the dressed mass $\overline m(g,T,\mu)$ may be more transparent 
from considering the above expressions in the high temperature approximation (and $\mu =0$ to simplify), upon using 
well-known $T\gg m$ limits of the thermal integrals \cite{laine} $J_1, J_2$, given  in 
Appendix \ref{ApphT}.
This gives from Eq.(\ref{mopd0}) 
\begin{equation}
{\overline m}^2(g,T,\mu=0) = T^2 \frac{\pi^2}{3} \left [ \frac{1}{2b_0 g} - 
\ln \left ( \frac{M e^{\gamma_E}}{\pi T } \right )  -8\pi^2 s_1  \right ]^{-1}\; \simeq \frac{3}{8} g\,T^2 +{\cal O}(g^2) ,
\label{mbar1L}
\end{equation}
or, equivalently using Eq.(\ref{g1L})
\begin{equation}
{\overline m}^2(T,\mu=0) = T^2 \frac{\pi^2}{3} \left [  \ln \left ( \frac{\pi T }
{ e^{\gamma_E -\frac{53}{84}}\,\Lambda_{\overline {\rm MS}}} \right ) \right ]^{-1}\;,
\label{mbar1Lexp}
\end{equation}
where we used $8\pi^2 s_1= -53/84$ for $N_f=3$.
As seen in Eq.(\ref{mbar1L}), for small coupling $\overline m$ admits a perturbative expansion 
having the expected form of a thermal screening mass. We stress however that $\overline m$
is unrelated to the perturbative Debye mass \cite{kapusta-gale},
which at one-loop order has the well-known expression (for $\mu=0$): 
\begin{equation}
m_{PT}^2 = \frac{g}{6}\,T^2 +{\cal O}(g^2) .
\label{mPT}
\end{equation}
In contrast $\overline m$ in Eq.(\ref{mbar1L}) represents an intermediate variational quantity, 
 whose meaning is merely once being inserted in $P({\overline m},g,T,\mu)$ 
to define the (optimized) physical pressure at a given order. 
Remark that, upon embedding RG invariance properties via the subtraction terms in Eq.(\ref{PRGPT}),
leading to $\overline m$ in Eq.(\ref{mopd0}), the LO RGOPT 
pressure (\ref{Prgoptd0simp}) involves nontrivial interaction terms. 
 Indeed upon perturbatively reexpanding Eq.(\ref{Prgoptd0simp}) using Eq.(\ref{mbar1L}), it  
can be seen to resum arbitrary higher order contributions, although only
those contributions induced by the specific leading order RG dependence\footnote{In the simpler
$O(N)$ $\phi^4$ model, the analogous LO RGOPT\cite{prdphi4} 
resums {\em all} large-$N$ contributions, reproducing the exactly known large-$N$ pressure\cite{phi4N}, 
 including nonanalytic terms $\sim \lambda^{3p/2}$, $p\ge 1$, typical of a boson gas pressure. }. 
Accordingly at LO and in the high-$T$
approximation, using Eq.(\ref{g1L}), Eq.(\ref{Prgoptd0simp}) takes the simpler form
\begin{equation}
\frac{P^{RGOPT}_0}{P_{SB}} \simeq 1-\frac{5}{14} \left [  \ln \left ( \frac{\pi T }
{ e^{\gamma_E -\frac{53}{84}}\,\Lambda_{\overline {\rm MS}}} \right ) \right ]^{-1} \;,
\label{PRGOPT1LPSb}
\end{equation}
normalized to the SB ideal quark gas $P_{SB}$ Eq.(\ref{Pqsb}) (here for $\mu=0$). 
The fact that the higher order contributions may be absorbed essentially into a one-loop running coupling
(for $\mu= 0$ and high-$T$ limits) is a peculiar LO feature of our construction:
as we will see below at NLO the more involved 
RG-induced higher order corrections are not so simply incorporated.
Another RGOPT feature is manifest in Eq.(\ref{PRGOPT1LPSb}): at high-$T$
the explicit $M$-dependence in Eq.(\ref{mbar1L}) has been automatically 
traded for a dependence in 
$g(\sim \pi T/\Lambda_{\overline {\rm MS}})$, consequently from scale invariance, 
rather than being an extra convenient scale choice 
$M\sim \pi T$ to absorb $\ln(M/\pi T)$ terms like in more 
standard (non-resummed) thermal perturbative expansions. 

The LO pressure Eq.(\ref{Prgoptd0simp}) is however not expected to give a very realistic
approximation of the complete higher order pressure,
as it only relies on LO RG-invariance properties embedded 
within an essentially free gas pressure. The LO dressed mass $\overline m$ of Eq.(\ref{mopd0}) with exact $T$-dependence  
is illustrated as function of $T$ in Fig. \ref{mbarmu0} (where it is also compared to NLO RGOPT
dressed masses to be specified below). The corresponding pressure Eq.(\ref{Prgoptd0simp}) is illustrated
e.g. in Figs.\ref{Pmg2vsPT} and \ref{Pmu0band}  for $\mu=0$ 
or in Figs. \ref{PMOPmu400} and \ref{PRGmu400} for $\mu\ne 0$, where it is also compared with NLO RGOPT and 
other NLO results.
We will next proceed to the more realistic NLO RGOPT pressure: most of the above features
will be maintained, except that the scale invariance can only be achieved approximately beyond LO, as we will examine. 
\subsection{Two-loop RGOPT}\label{sec2L}
At NLO the RG Eq.(\ref{RGred}) is no longer automatically satisfied by Eq.(\ref{P2LRGOPT})
with (\ref{aQCD}),
and can be thus considered as an independent constraint.
Following \cite{prlphi4,nlsm,prdCOLD} we can in principle use  
either the MOP Eq.(\ref{mop}) {\em or} the RG Eq.(\ref{RGred}), defining two possible alternative
dressed mass $\overline m(g,T,\mu)$: we will consider in the following both prescriptions, for completeness 
and comparison purposes. Accordingly the coupling $g(M)$ is simply 
determined from standard PT, i.e. with its running at
(exact) two-loop order given by Eq.(\ref{g2L}) and scale $M$ chosen as 
a combination of $\pi T$ and $\mu$ when both $T, \mu$ are non-zero. 
 As shown above in Sec.~\ref{sec1L}, the LO RGOPT pressure exhibits one-loop-exact scale invariance as a consequence of the simple 
form of Eq.(\ref{P1LRGOPT}) only depending on $b_0$. While at NLO or beyond, $P^{RGOPT}(\overline m(g),M,\cdots)$ inevitably 
has a remnant scale dependence:
the basic reason is that the subtractions in Eq.(\ref{PRGPT}) solely guarantee RG invariance up to
remnant higher order terms, $\sim m^4 g^2$ at NLO. This feature cannot be drastically reduced by the subsequent variational modification 
in Eq.(\ref{Lint}). Concretely the exact NLO running coupling Eq.(\ref{g2L}), that depends only on $b_i$ coefficients in 
Eqs.(\ref{b0def}),(\ref{b1def}), cannot perfectly cancel the explicit scale dependence of Eq.(\ref{P2LRGOPT}), that also involves anomalous
mass dimension coefficients $\gamma_i$ in (\ref{gamdef}) reminiscent of the originally NLO massive theory. 
Nevertheless it is a nontrivial consequence of our above construction through steps 1)--4) in Sec.\ref{secRGOPT}, preserving RG invariance 
at least at the same perturbative level as Eq.(\ref{RGeq}), that the remnant scale dependence remains formally of higher order $\sim m^4 g^2$.
 Accordingly this NLO RGOPT scale dependence, that we will exhibit by varying the scale by a factor 2 around central $M\sim 2\pi T$ (for $\mu=0$), 
 is generically milder\cite{prlphi4,prdphi4,nlsm} than for standard PT and HTLpt. 
 It is thus expected to remain moderate (and to further decrease at NNLO) 
 even for relatively low temperature where the resulting dressed thermal mass is not necessarily
 perturbatively screened.
 Using the standard PT running coupling also more directly compares with the same common prescription in 
 other related thermal resummations approaches, like HTLpt typically. But one should keep in mind 
 that identifying the arbitrary renormalization scale $M$ to be ${\cal O}(\pi T)$ is strictly valid only at 
 sufficiently high temperatures.

  As mentioned above, Eq.(\ref{aQCD}) has the property to select a unique NLO solution matching the perturbative asymptotic
freedom (AF) behavior for $g\to 0$ at $T=0$. 
As it happens however regarding the NLO quark pressure Eq.(\ref{P2LRGOPT}),  
imposing either Eq.(\ref{mop}) or Eq.(\ref{RGred}) both fail to readily give a real dressed mass
$\overline m(g,T,\mu)$ for a substantial part of the physically relevant $T,\mu$ range. This is 
admittedly a technical burden of such methods, but the occurrence of complex variational solutions has no deeper physical meaning.
Rather, it may be viewed to some extent as an accident of the specific $\ms$ scheme in which the
original perturbative coefficients were calculated, given that nonreal solutions are often 
expected upon exactly solving nonlinear equations, like in the present case solving for $m$ 
the NLO Eqs.(\ref{mop}) or (\ref{RGred}). At the same time we wish 
to maintain these relations as exact as possible in order to capture RG resummation properties beyond PT. 
A crude escape could be simply to take the real part of the solutions, 
but that potentially loses some of the sought RG properties.
The nonreal solution issue also occurred in the simpler $T=\mu=0$ case \cite{JLalphas} as well as within 
the $T=0, \mu\ne 0$ cold quark matter application \cite {prdCOLD}, 
where it was  cured by performing a renormalization scheme change (RSC)\cite{JLalphas}.
The latter allows for the  recovery of  real solutions by modifying perturbative coefficients while 
keeping RG consistency by definition. Of course
for such a solution to work the RSC should not be arbitrary, but
fixed by a sensible prescription, and importantly such that it remains a moderate (i.e. perturbative) deviation from
the original scheme. 
More specifically in \cite {prdCOLD} a relevant NLO RSC parameter $B_2$ was uniquely fixed by requiring collinearity
of the RG and MOP curves in the $\{m, g\}$ plane (that precisely expresses the recovering of real solutions). 
Technically this implies to nullify the determinant of partial derivatives of the RG and MOP equations, 
and to solve the latter together with, e.g., Eq.(\ref{mop}) for $\{B_2, \overline m(B_2,g)\}$. While solving such a
coupled system was easily manageable for the (entirely analytical) $T=0, \mu \ne 0$ NLO expressions in \cite{prdCOLD}, 
it becomes numerically quite challenging for the rather involved $T,\mu \ne 0$ NLO dependence from Eq.(\ref{P2LRGOPT}).
Therefore in the present study, seeking as much as possible for simplicity, 
we will exploit the RSC arbitrariness quite similarly to recover real solutions, 
but via simpler alternative prescriptions precisely defined in next Sec. \ref{secpresc}.
The reader mainly interested in concrete results for the thermodynamical quantities may skip
this section proceeding directly to Sec.\ref{secnum}. 
\section{NLO prescriptions}\label{secpresc}
\subsection{Simple RSC parametrization}
Let us first specify for our later purposes the RSC to be used.
Since one basically introduces a variational (quark) mass, the most natural and simplest RSC can 
be defined by modifying only the mass parameter:
\begin{equation}
m \to m^\prime ( 1+ B_2 g^2) \,,
\label{RSC}
\end{equation}
where a single $B_2$ coefficient parametrizes a perturbative NLO scheme change
from the original ${\overline{\text{MS}}}$-scheme 
\footnote{Eq.(\ref{RSC}) has also the welcome property that it does not affect the definition of 
the reference QCD scale $\Lambda_{\overline{\text{MS}}}$, in contrast with a similar perturbative 
modification acting on the coupling (see Ref.\cite{JLalphas} for details).}.  
As is well-known, for a perturbative series truncated at
order $g^k$ (like in the present case the original order-$g$ pressure Eq.(\ref{PPT})), 
different schemes differ formally by
remnant terms of order ${\cal O}(g^{k+1})$, such that the difference between two schemes 
is expected to decrease at higher orders for sufficiently small coupling value. 
 Note that we perform the perturbative RSC Eq.(\ref{RSC}) consistently on the original PT
expression (\ref{PPT}) {\em prior} to its
modification induced from Eq.(\ref{Lint})
with the subsequent $\delta$-expansion.
The net RSC modification to the pressure 
is to add an extra term, $-4g\, (m^\prime)^4 s_0 B_2 $, entering thus the resulting
exact NLO Eq.(\ref{mop}) or Eq.(\ref{RGred}). Thus
Eq.(\ref{RSC}) modifies the latter equations purposefully, now considering those equations as 
constraints for the arbitrary mass $m^\prime$, after the 
modifications from Eq.(\ref{Lint})\footnote{To avoid excessive notation proliferation, 
in what follows once having performed the replacement implied by Eq. (\ref{RSC}) we simply rename $m^\prime \to m$ the 
variational mass to be determined from 
Eq.(\ref{mop}) or Eq.(\ref{RGred}).}. 
Accordingly $B_2$ may be considered as an extra variational parameter, quite similarly to $m$, 
thus to be fixed by a definite prescription as will be specified below.
\subsection{AF-compatible NLO dressed mass solutions}
To identify some relevant properties of the sought dressed $\overline m(g,T,\mu)$ solutions 
we consider first the MOP Eq.(\ref{mop}) more explicitly at NLO, thus applied to Eq.(\ref{P2LRGOPT}). 
It is convenient to formally solve it in a first stage for $\ln [m^2/M^2]$, as that would give simply an
exact quadratic equation at $T=\mu=0$. Accordingly 
the two equations (that are implicit in $m$ for $T, \mu \ne 0$) can be conveniently written,  
after straightforward algebra, as
\begin{equation}
 -\ln \frac{ m^2}{M^2} + B_{mop} 
 \mp \frac{2\pi}{3 g} \sqrt{D_{mop}} = 0 \; , 
 \label{OPTexact}
\end{equation}
where for $T, \mu \ne 0$,  $B_{mop}$ and $D_{mop}$ take a relatively compact form:
\begin{equation}
 B_{mop} = - \frac{7\pi^2}{9 g} +\frac{5}{6}  +4\pi^2 \left (J_2^\prime+\frac{T^2}{m^2} J_2 \right ) ,
 \label{Bopt}
\end{equation}
\begin{eqnarray}
D_{mop} &= &  9\frac{\pi^2}{4} -\frac{47}{6} g  - g^2 \left (\frac{35}{16\pi^2} + 288\frac{\pi^2}{7} B_2 \right )  \nonumber \\
&& +36 \pi^2 g^2 \left(J_2^{\prime\,2}+ \frac{T^4}{m^4} J_2^2 \right)
 + 9 g  (g -2 \pi^2) \left(\frac{T^2}{ m^2} J_2 -J_2^\prime \right)   \nonumber \\
&& +8 \pi^2 g^2 \frac{T^2}{ m^2} \left(3J_2-1\right)J_2^\prime  
 - 48 \pi^2 g^2 \left(J_3^\prime + \frac{T^2}{ m^2} J_3 \right)  \; ,
\label{Delta}
\end{eqnarray}
where $J_i\equiv J_i(m^2/T^2,\mu/T)$, $J_i^\prime \equiv \partial_x J_i(x)$, (note that here $x\equiv m^2/T^2$).
In Eq.(\ref{Delta})
we explicitly separated the $T, \mu$-independent part within $D_{mop}$ in the very first line to make its 
$T, \mu \to 0$ limit clear
(remark also that $D_{mop}(T=0)$ does not depend on $m$).\\
One first property of Eq.(\ref{OPTexact}) is exhibited 
from expanding it perturbatively to the first few terms. That gives
\begin{equation}
\ln \frac{\overline m^2}{M^2}(-) \simeq -\frac{16\pi^2}{9 g} +\frac{139}{54} +8\pi^2 \frac{T^2}{\overline m^2} J_2 +{\cal O}(g) ,
\label{mAFT0}
\end{equation}
and
\begin{equation}
\ln \frac{\overline m^2}{M^2}(+) \simeq \frac{2\pi^2}{9 g} -\frac{49}{54} +8\pi^2  J_2^\prime +{\cal O}(g) .
\label{mnAFT0}
\end{equation}
One easily recognizes that, for $T\to 0$ the leading term for $g\to 0$ in $\overline m^2(-)$ has the correct 
AF behavior: 
$\ln \frac{\overline m^2}{M^2}(-)\sim -1/(b_0 g)$, noting that 
$b_0=9/(16\pi^2)$ (for $N_f=3$), which as recalled above is a compelling requirement of the RGOPT.
In contrast the other $(+)$ solution has a wrong sign and coefficient, thus
drastically in contradiction with
AF for $g\to 0$.  
Therefore clearly only the above equation (\ref{OPTexact}) with $(-)$ is to be selected. 
It is further instructive to investigate the behavior of those two solutions for $T\ne 0$, 
taking for simplicity the high-$T$
approximation (and $\mu=0$, see
Eq.(\ref{J2hT})). After straightforward algebra
one obtains, for the first few perturbative expansion terms: 
\begin{equation}
\frac{\overline m^2_{(-)}}{T^2} =  \frac{3}{8} g \left[1-\frac{3}{8\pi^2}g \left (3 L_T +\frac{85}{36} \right ) \right]^{-1} + g^2\,
\left (\frac{67}{288\pi^2} +6 J_3(0,0) \right) +{\cal O}(g^3) ,
\label{mAFhiT}
\end{equation}
where we defined for short
\be
L_T\equiv \ln \left(\frac{M e^{\gamma_E}}{\pi T}\right) .
\ee
As seen the AF-compatible solution $\overline m(-)$ has a typical perturbative 
thermal screening mass behavior $m \sim \sqrt{g}\, T$, with a coefficient here mainly determined by RG 
properties (notice that the first order term is consistent with our LO above 
result, Eq.(\ref{mbar1L})).
In contrast 
the non-AF-compatible Eq.(\ref{OPTexact}) with $(+)$ has $\overline m(+)$ solutions for $T,\mu \ne 0$  
having a coupling dependence that cannot be cast into the form of a perturbative expansion for small enough $g$. 
Moreover the corresponding real solutions generally give $m/T\gg 1$, unless $g$ is very large
(see Appendix \ref{Appnumsol}).
The latter features give further compelling reasons to rejecting this non-AF solution also for the $T,\mu \ne 0$ case.
Thus as anticipated the AF-compatibility criterion leads to a unique MOP solution.\\

\begin{figure}[h!]
\centerline{ \epsfig{file=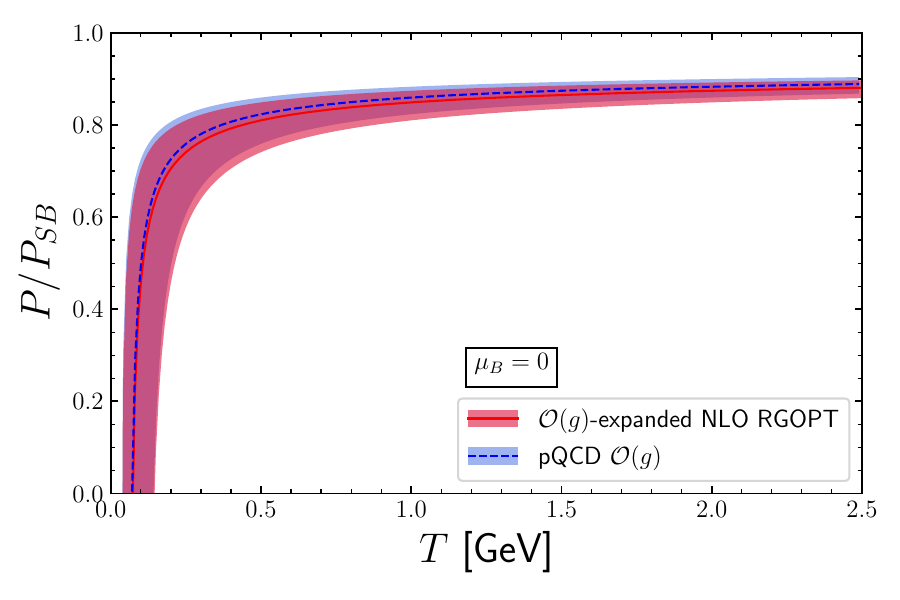,width=0.5\linewidth,angle=0}}
\caption{Perturbatively re-expanded NLO RGOPT pressure $P(T,\mu=0)$ (red band) compared with standard perturbative 
NLO pressure  Eq.(\ref{PPTm0}) (blue band), with scale dependence  $\pi T\le M\le 4\pi T$. }
\label{PrgoptPTvsPT}
\end{figure}

The purely perturbative expansion Eq.(\ref{mAFhiT}) is however not expected to give 
a very good approximation for relatively low~\footnote{In particular the exact NLO $\overline m$ as 
obtained below can be such that $\overline m/T >1$ at sufficiently low $T$ (see Fig. \ref{mbarmu0}), 
somewhat invalidating the high-$T$ approximation.} $T$, and obviously not useful anyway for $\mu\ne 0$. 
Before we proceed below with the more elaborate RSC 
prescription to solve exactly Eq.(\ref{OPTexact}), it may be
instructive to illustrate the results of using the simple perturbative solution Eq.(\ref{mAFhiT}), 
inserted in our NLO pressure
Eq.(\ref{P2LRGOPT}), and with the resulting expression being truncated simply at first order in $g$ (i.e. this is 
accordingly the NLO generalization of Eq.(\ref{PRGOPT1LPSb})).
This is shown in Fig.~\ref{PrgoptPTvsPT}, compared to the true standard NLO 
massless quark PT pressure Eq.(\ref{PPTm0}). 
This result represents a good consistency check of our procedure: the two pressures are not strictly identical 
but very close, since after expressing the optimized mass 
$\overline m(g)$, the RGOPT is expected to approximate a massless theory.
(Note that replacing $m$ in Eq.(\ref{P2LRGOPT}) instead by, e.g.,    
the standard thermal Debye mass Eq.(\ref{mPT}), would give results more drastically departing 
from the massless PT pressure). Now more interestingly, the main purpose of the RGOPT is rather 
to provide higher order deviations from standard PT, induced by higher order RG-induced terms, 
as we will exhibit next. 
\subsection{NLO mass optimization prescription}\label{secMOP}
Going back to the exact MOP Eq.(\ref{OPTexact}), Eq.(\ref{Delta})
involves the RSC parameter $B_2$
as induced from Eq.(\ref{RSC}). In the original $\ms$ scheme,
i.e. $B_2\equiv 0$, $D_{mop}$ from Eq.(\ref{Delta}) can take negative values for not particularly large couplings\footnote{
For example for $T=\mu=0$ where only the first three terms of Eq.(\ref{Delta}) are nonvanishing,
$D_{mop} \le 0$ for rather moderate $g \ge 2.64$, i.e $\alpha_S\ge 0.21 $.}.
As anticipated above it therefore renders the (exact) $\overline m(g,T,\mu)$ solution not always real, 
except in a rather limited 
range of physically interesting $T$ and/or
$\mu$ values. Remark however that since the (perturbatively leading) first term in 
Eq.(\ref{Delta}) is positive, this loss of real $\overline m$ solutions arises 
solely when considering the exact Eq.(\ref{OPTexact}): 
now since all our results were obtained from modifying perturbative NLO expressions,
one may simply  expand perturbatively Eq.(\ref{OPTexact}), obtaining therefore a real expression at arbitrary orders 
(as partially illustrated by the first few orders of such an expansion in Eq.(\ref{mAFhiT})). But 
it is soon realized that this is a poor approximation of the actual exact expression,  
even for $g$ slightly below the value at which $D_{mop}$ becomes negative. Accordingly  such perturbative expansion would 
partly lose the sought RG properties, due to RG-consistent contributions being perturbatively truncated.   
Now with $D_{mop}$ not too far from being positive, a more efficient way to recover real solutions is from an appropriately
chosen $B_2$ value such that $D_{mop}>0$. 

\begin{figure}[h!]
\centerline{ \epsfig{file=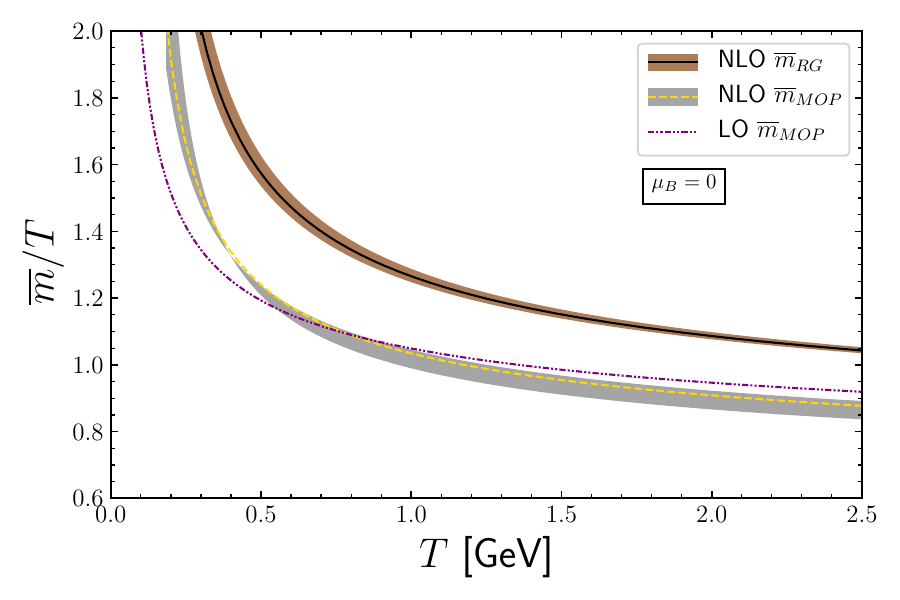,width=0.5\linewidth,angle=0}}
\caption{Exact LO RGOPT thermal mass (dot-dashed) compared with exact MOP and RG NLO 
thermal mass for $\pi \le M \le 4\pi T$ at $\mu_B=0$. }
\label{mbarmu0}
\end{figure}

\begin{figure}[h!]
\centerline{ \epsfig{file=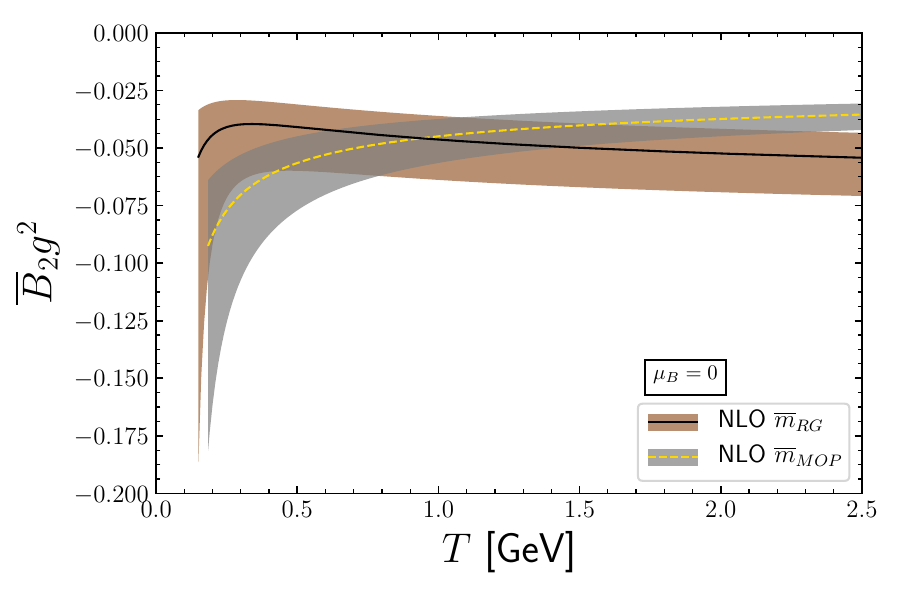,width=0.5\linewidth,angle=0}}
\caption{RSC parameter $B_2 g^2(M)$ for the MOP and RG prescriptions for $\pi T \le M \le 4\pi T$ at $\mu_B=0$. }
\label{B2g2mu0}
\end{figure}

Let us thus define precisely our 
prescription for the MOP Eq.(\ref{OPTexact}):
in a first stage we fix the arbitrary RSC parameter $B_2$ in Eq.(\ref{Delta}) such that $D_{mop}>0$.
Next, the resulting modified AF-matching Eq.(\ref{OPTexact}) with $(-)$ is solved exactly 
(numerically) for $\overline m(g,T,\mu)$, recovering real solutions for practically most 
relevant $g$ values. 
 Note that simply requiring $D_{mop} \ge 0$ does not give a unique prescription, but it happens to 
be rather constrained: first $D_{mop}=0$ is excluded, as it would spoil the crucial AF-compatibility
of Eq.(\ref{mAFT0}), that at least requires the LO (first) term of Eq.(\ref{Delta}). On the other hand
if $D_{mop}>0 $ would be too large, the AF-matching $(-)$ Eq.(\ref{OPTexact}) 
would take too negative values no longer giving a real solution (i.e., it cannot cross the $x$-axis).
Since the problem comes from some negative terms within Eq.(\ref{Delta}), a 
prescription that appears minimal is to fix $B_2$ such as to cancel 
solely the largest (in magnitude) $T,\mu$-independent negative term within Eq.(\ref{Delta}),
$-(47/6)g$. Explicitly that gives:
\begin{equation}
 B_2= -\frac{329}{1728\pi^2\,g} \; .
 \label{B2simp}
\end{equation}
The latter $B_2$ prescription is very simple, and the resulting $\overline m_{MOP}$ solution 
remains real for practically all 
physically relevant $g(T,\mu)$ values, while still including nontrivial higher order 
corrections induced from all remnant  terms of  Eq. (\ref{OPTexact}).
Other slightly different $B_2$-fixing prescriptions are possible for $T,\mu \ne 0$, but 
a notable property is that for different  $B_2$ choices, that imply
different exact $\overline m(B_2)$ solutions,
 the resulting physical pressure $P(\overline m(B_2),B_2,\cdots)$ 
Eq.(\ref{P2LRGOPT}) happens to be 
largely insensitive to those unphysical $B_2$ parameter choices provided that $\overline m(B_2)$ remains real. 
 This welcome feature is to be traced to the underlying RSC properties, together with the further  
perturbative screening from Eq.(\ref{mAFhiT}): ${\overline m^2} \sim (3/8) g T^2+{\cal O}(g^2)$: as easily checked, 
$B_2$ from Eq.(\ref{RSC}) 
only appears at higher order ${\cal O}(g^3)$ both in 
 the perturbatively expanded $\overline m$ Eq.(\ref{mAFhiT}) 
 and corresponding re-expanded pressure from Eq.(\ref{P2LRGOPT}). 
In other words once $B_2$ is adjusted
to recover a real $\overline m$ solution of Eq.(\ref{OPTexact}), 
the discrepancies between possibly different $B_2$ 
prescriptions are somewhat hidden within perturbatively higher order terms. \\
To close this subsection, 
we illustrate in Fig. \ref{mbarmu0} the resulting dressed thermal masses as function of the temperature,
both at LO from Eq.(\ref{mopd0}), and NLO from MOP Eqs.(\ref{OPTexact}), (\ref{B2simp}).
 As already mentioned their behavior is essentially that of screening thermal masses, 
except that those are determined from RG properties. We also compare  in Fig. \ref{mbarmu0} with
the similar dressed thermal mass as obtained from the  alternative RG prescription, 
giving Eqs.(\ref{RGexact}) with (\ref{B2sq0}), as we will specify in next subsection.
%%%%%%
%%%%%%%%
Correspondingly Fig. \ref{B2g2mu0} illustrates the relevant RSC deviation  $B_2 g^2$ in Eq.(\ref{RSC})
resulting from Eq.(\ref{B2simp}) as function of $T$. As an important crosscheck, 
it shows that the departure from the original $\ms$-scheme 
remains quite moderate.
%%%%%%%%
%
\subsection{Alternative NLO RG prescription}\label{secRG}
Alternatively, the other very relevant prescription, as anticipated in Sec.\ref{sec2L}, is to consider the RG Eq.(\ref{RGred}) instead of the MOP 
Eq.(\ref{OPTexact})  
to determine the dressed mass $\overline m(g,T,\mu)$. Once expressed
for $\ln (m^2/M^2)$ it takes a similar quadratic form as Eq.(\ref{OPTexact}), conveniently normalized as
\begin{equation}
 -\ln \frac{ m^2}{M^2} + B_{rg}
 \mp \frac{8\pi^2}{g} \sqrt{\frac{2}{3} D_{rg}} = 0 \; , 
 \label{RGexact}
\end{equation}
where explicitly
\begin{equation}
 B_{rg} = 
 -\frac{1}{b_0\,g} +\frac{172}{81} -\frac{64}{81} \left (\frac{4g}{9\pi^2}\right )\,\frac{1}
 {1+\frac{4g}{9\pi^2}}
 +8\pi^2 \frac{T^2}{m^2} J_2 ,
 \label{Brg}
\end{equation}
and
\begin{equation}
D_{rg} = -\left (\frac{3}{7} B_2  +\frac{11}{384\pi^4}\right )g^2 -
\frac{g}{27}\frac{(4 g + 81 \pi^2)}{ (4g + 9\pi^2)^2} + g^2\,
 \frac{T^4}{m^4} J_2 \left (J_2 -\frac{1}{6} \right )  - g^2\,\frac{T^2}{m^2} J_3 .
 \label{Drg}
 \end{equation}
Now, similarly to the previous MOP Eq.(\ref{OPTexact}), for $B_2=0$ one obtains generally nonreal solutions since 
in $D_{rg} $ some contributions 
happen to be negative.  In contrast with Eq.(\ref{OPTexact}) however,  
the crucial AF-matching for the RG solution is already guaranteed solely from the first term in (\ref{Brg}), 
up to higher order terms. These features strongly suggest the prescription
fixing the arbitrary RSC parameter $B_2$ as simply to fully cancel $D_{rg}$:
\begin{equation}
D_{rg}(B_2) \equiv 0 .
 \label{B2sq0}
 \end{equation}
Eq.(\ref{B2sq0}) determines $B_2$ trivially using Eq.(\ref{Drg}), 
leading to a single real AF-compatible solution $\overline m_{RG}$ determined from the first two terms of Eq.(\ref{RGexact}),  
the latter being still an implicit equation in $m$ for $T,\mu\ne 0$
via $J_2$ entering Eq.(\ref{Brg}).
 Eq.(\ref{B2sq0}) may appear a rather peculiar choice, 
but there happen to be very few other choices to recover a real RG solution. We stress
that for any (MOP or RG) prescriptions the resulting $\overline m(B_2)$ is an intermediate variational parameter 
without much physical meaning outside its use in the pressure. Here the resulting 
$\overline m_{RG}(B_2)$ still involves arbitrary higher order contributions, 
as well as nontrivial $T,\mu$ dependence via  
$B_{rg}$ in Eq.(\ref{Brg}).
Similarly as for the MOP above prescription, 
we have checked that for other $B_2$ choices, as long as being moderately different from Eq.(\ref{B2sq0}),
our numerical RG results for $T, \mu \ne 0$ are not strongly 
dependent upon those choices.

 The dressed exact thermal mass $\overline m_{RG}$ resulting from Eqs.(\ref{RGexact}),(\ref{B2sq0}) 
 is illustrated
as function of the temperature in Fig. \ref{mbarmu0}, and compared with the previously discussed 
LO mass from Eq.(\ref{mopd0}) and $\overline m_{MOP}$ from Eqs.(\ref{OPTexact}), (\ref{B2simp}). 
As seen the dressed masses are numerically quite different, but such differences in the 
two alternative NLO variational masses are drastically reduced within the physical pressure
as will be illustrated below.
The corresponding RSC deviation $B_2 g^2$ obtained from Eq.(\ref{B2sq0}) is illustrated in Fig. \ref{B2g2mu0}
as function of $T$, and compared to the similar MOP $B_2 g^2$ from Eq.(\ref{B2simp}). 
Notice that despite the visible discrepancies between the two expressions, they are numerically not drastically
different and both behave smoothly, except at very low $T\lesssim 0.5$ GeV: as already mentioned above the important
feature is that the induced departure from the original $\ms$-scheme 
remains moderate.
\section{RGOPT pressure results at NLO}\label{secnum}
 To obtain the full benefit from the  RGOPT, in particular the optimally reduced
scale dependence, a price to pay as a result of the variational approach 
is to first solve exactly numerically for the dressed mass 
(either from Eq.(\ref{mop}) or alternatively Eq.(\ref{RGred})), prior to its
use in the RGOPT pressure at NLO,  Eq.(\ref{P2LRGOPT}). 
Such a procedure is moreover 
complicated by the onus of complex solutions, cured by the appropriate RSC
as specified above in Sec. \ref{secpresc}.
But the relevant NLO expressions (\ref{OPTexact}) or alternatively (\ref{RGexact}) are reasonably simple and 
the numerical procedure is straightforward.  
 Before illustrating the resulting exact NLO RGOPT pressure, 
we start this section with another intermediate (more perturbative) prescription, 
to show the gradual improvement typically concerning the remnant scale dependence.
\subsection{A simple perturbative approximation}\label{simpapp}
The simplest we can do to recover 
real solutions without going through RSC considerations as elaborated on previously in Sec.\ref{secpresc},  
while capturing at the same time more accurate $T,\mu$ dependence,
is to expand $\overline m$ from the MOP Eq.(\ref{OPTexact}) perturbatively to NLO ${\cal O}(g^2)$, but keeping
the exact thermal integrals in the resulting expression. This gives after simple 
algebra~\footnote{As an algebraic subtlety, one should {\em first} expand perturbatively the (AF-matching) 
Eq.(\ref{OPTexact}) with $(-)$ before
to solve it formally for $m^2/T^2$, otherwise one loses the latter AF-matching properties.}
\begin{equation}
\frac{m^2_{MOP}}{T^2}  = 
 \frac{9}{2} g J_2 + 
 g^2 \left[ \frac{17}{9} J^\prime_2 (1-12 J_2)+
\frac{34}{3} J_3
+\left (\frac{20371}{1728\pi^2} -\frac{81}{32\pi^2} \ln \frac{m^2}{M^2} \right ) J_2 \right] ,
\label{mbarg2}
\end{equation}
therefore still to be solved numerically as an implicit function
since $J_i\equiv J_i(\frac{m}{T},\frac{\mu}{T})$. 
\begin{figure}[h!]
\centerline{ \epsfig{file=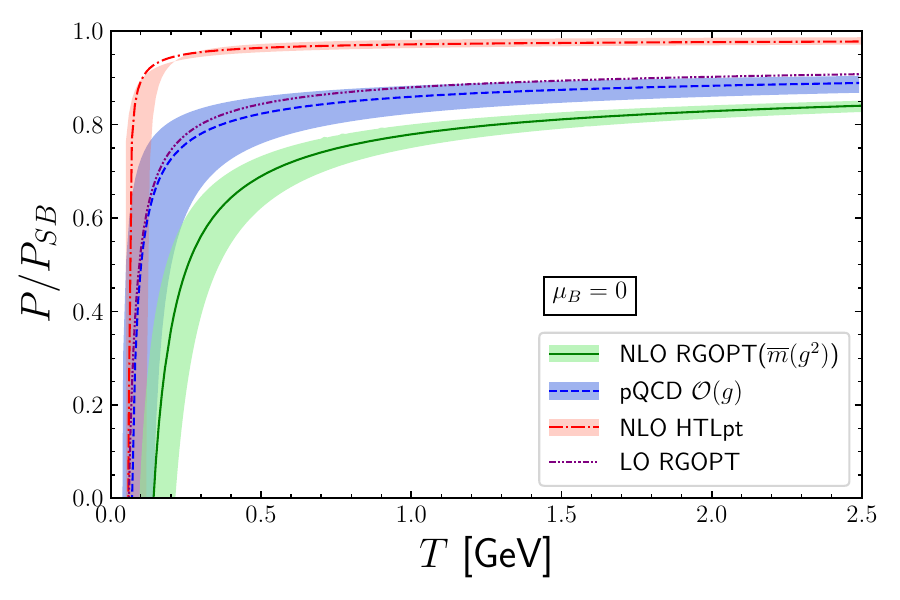,width=0.5\linewidth,angle=0}}
\caption{Comparison of NLO RGOPT quark pressure Eq.(\ref{P2LRGOPT}) with $\overline m(g^2)$ (green, thin lines), 
LO RGOPT (dotdashed), NLO PT (blue, dashed), NLO HTLpt quark pressure (red, dotted) 
with scale dependence $\pi T\le M \le 4\pi T$ (bands) and central scale $M=2\pi T$ (lines) at $\mu_B=0$.}
\label{Pmg2vsPT}
\end{figure}
The above expression readily gives a real solution, and 
allows to consider $\mu\ne 0$ within the thermal integrals (and within
the running coupling as well) while still keeping a relatively simple ``perturbative-like'' expression. 
Inserting the solution of Eq.(\ref{mbarg2})
into the RGOPT NLO quark pressure Eq.(\ref{P2LRGOPT}) (keeping also 
exact thermal integrals consistently in the latter), gives the results illustrated for $\mu=0$ in Fig. \ref{Pmg2vsPT}, 
compared with the standard NLO PT pressure Eq.(\ref{PPTm0}), and also with the NLO HTLpt (quark) pressure.  
 (NB for a consistent comparison with the latter at this stage,
we have extracted the sole quark contributions within the complete QCD NLO HTLpt pressure,
which is not a trivial separation as in the case of NLO pQCD. How to do this precisely is explained in Appendix \ref{AppHTL}).

 Alternatively, proceeding similarly with the RG Eq.(\ref{RGexact}) and (\ref{B2sq0}) gives
\begin{equation}
\frac{m^2_{RG}}{T^2}  = 
 \frac{9}{2} g J_2 + 
 \frac{g^2}{32\pi^2} \, \left( 172 -81 \ln \frac{m^2}{M^2}  \right) J_2,
\label{mbarRGg2}
\end{equation}
observing that the LO term and the $\ln M$ dependence are
identical to those in Eq.(\ref{mbarg2}). This illustrates that
although the MOP and RG prescriptions are quite different if considering their exact determinations, 
perturbatively they differ only by ${\cal O}(g^2)$ terms, thus formally higher order than the original 
NLO perturbative pressure from which they were both constructed.
Moreover, inserting Eq.(\ref{mbarRGg2}) within Eq.(\ref{P2LRGOPT}) gives almost identical
results as in Fig. \ref{Pmg2vsPT}. Note also that in both Eqs.(\ref{mbarg2}) and (\ref{mbarRGg2}) the 
running $g(M)$ exactly cancels the $M$-dependence at ${\cal O}(g^2)$, as easily checked using Eqs.(\ref{g1L}),
(\ref{b0def}), and (\ref{J2hT}).

As seen in Fig.~\ref{Pmg2vsPT} the RGOPT pressure with the (MOP or RG) $\overline m(g^2)$
approximation has a more pronounced decrease, i.e. 
a departure from the ideal gas limit, than the standard NLO PT (pQCD) 
quark pressure and than LO RGOPT for moderate and low $T$ values,
that is mainly traced to the higher order $g^2$ contributions in Eq.(\ref{mbarg2}) or Eq.(\ref{mbarRGg2}). 
Actually, it is rather 
closer to the higher orders standard pQCD pressure, as will be illustrated below,
partly due to Eq.(\ref{P2LRGOPT}) and the thermal functions $J_i$ being kept exact. (If perturbatively
reexpanded, the resulting pressure gets back closer to the NLO pQCD result).
This is in contrast with the NLO HTLpt pressure,  
that remains very close to the ideal gas limit except at very low $T$ as seen in Fig.~\ref{Pmg2vsPT}~\footnote{We
mention that the NLO HTLpt pressure in Fig.~\ref{Pmg2vsPT} (and similarly below in Figs.~\ref{Pmu0band}-\ref{PRGpresc},
Figs.~\ref{PMOPmu400}-\ref{PRGmu400})
is somewhat different than the results in Ref.\cite{HTLptqcd2L}, specially at very low $T$. This is due to considering
here only its pure quark contributions, and partly also from using the exact Eq.(\ref{g2L})
instead of a more approximate two-loop running expression used in \cite{HTLptqcd2L}.}. 
In Fig. \ref{Pmg2vsPT} the RGOPT pressure also exhibits a better renormalization scale dependence 
as compared with NLO pQCD (at least for $T> 1$ GeV), 
although this is only a moderate improvement. Very similar results are obtained for $\mu \ne 0$, that
we omit to illustrate. We will see  below that the more elaborate untruncated 
RGOPT pressure, accounting for higher orders in $\overline m(g)$, has a more drastically improved scale dependence, 
which is a main expected RGOPT feature.
\subsection{Hot quark matter: $T\ne 0$, $\mu=0$}
\subsubsection{MOP prescription}
\begin{figure}[h!]
\centerline{ \epsfig{file=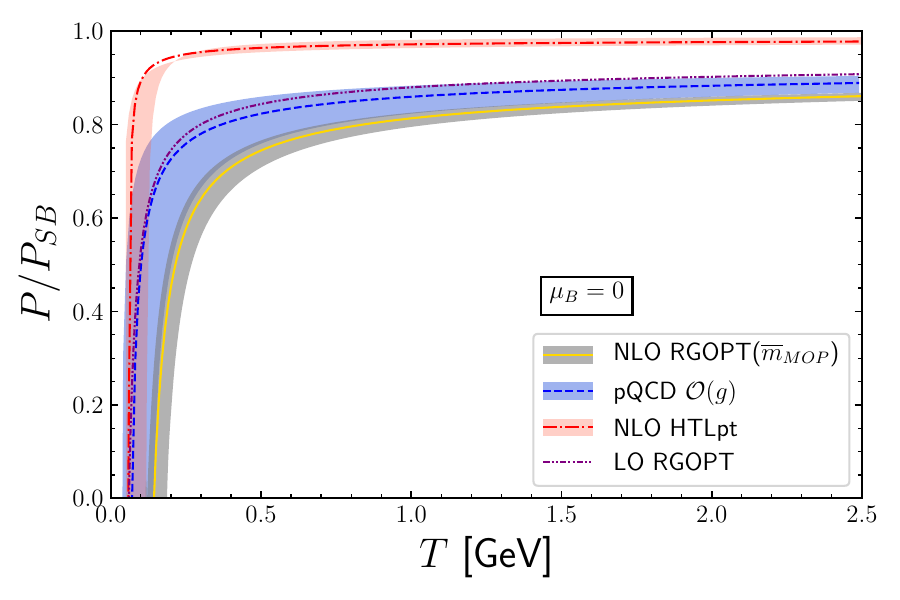,width=0.5\linewidth,angle=0}}
\caption{ RGOPT quark pressures as function of temperature 
at LO and NLO (MOP prescription) compared with standard NLO PT (pQCD) and NLO HTLpt pressures, 
with scale dependence $\pi T \le M \le 4\pi T$ at $\mu_B=0$. }
\label{Pmu0band}
\end{figure}

The resummation properties of the NLO RGOPT become more evident  
when one compares it with the standard perturbative one (pQCD) at the same NLO. 
We illustrate (first for $\mu=0$) the exact NLO RGOPT pressure $P(\overline m,g,T,\mu)$
obtained from our first $\overline m_{MOP}$ prescription, defined 
by solving Eqs.(\ref{OPTexact}),(\ref{B2simp}) (as explained in details Subsec.\ref{secMOP}). 
In Fig. \ref{Pmu0band}
the pressure is displayed as function of the temperature,
compared with the LO RGOPT and the standard NLO pQCD Eq.(\ref{PPTm0}), for the scale dependence $\pi T\le M\le 4\pi T$. 
The reduction of scale dependence stemming from the now exact (untruncated) NLO RGOPT appears substantial (about a factor $\sim 2$ 
improvement for e.g. $T\sim 1$ GeV). 
The HTLpt NLO (quark)
pressure\cite{HTLptqcd2L} is also shown in the same figure for comparison. 
 We observe that the (NLO) quark HTLpt pressure 
has a small residual scale dependence for most $T$ values (which is partly a consequence 
of limiting it to the quark only contribution), but does not depart very much from the ideal gas limit, 
in contrast with the RGOPT pressure. This latter feature is similar concerning the complete QCD NLO HTLpt\cite{HTLptqcd2L}),
while a more drastic departure from the ideal gas is only obtained 
at NNLO for HTLpt\cite{HTLptqcd3L}. 
\bigskip

%%%%%%%%%
\begin{figure}[h!]
    \centerline{ \epsfig{file=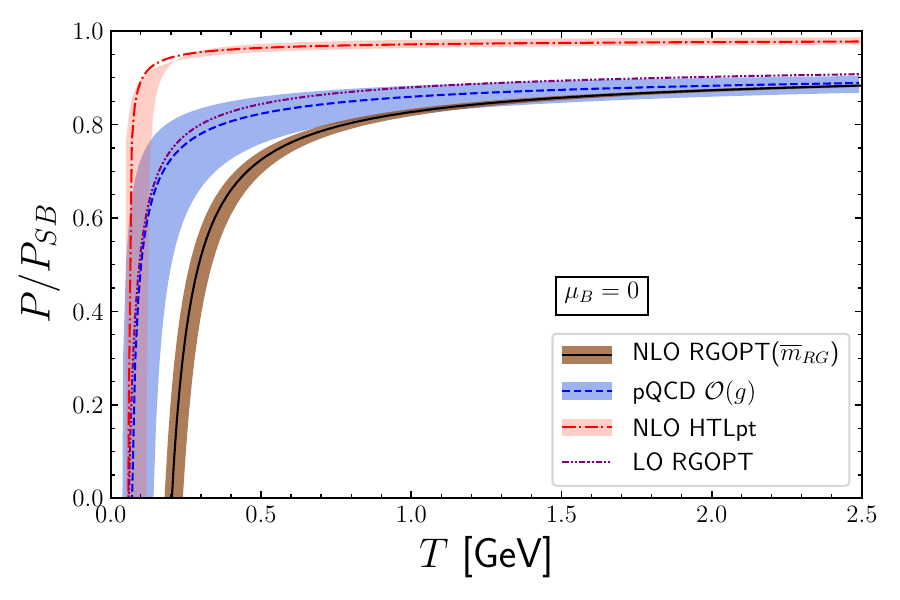,width=0.5\linewidth,angle=0}}
\caption{Same captions as for Fig.\ref{Pmu0band} but with the RGOPT pressure 
obtained from alternative $\overline m_{RG}$ prescription Eqs.(\ref{RGexact}), (\ref{B2sq0}).  }
\label{PRGpresc}
\end{figure}
%%%%%%%%%%%%%
%
\subsubsection{Alternative RG prescription}
\begin{figure}[h!]
\centerline{ \epsfig{file=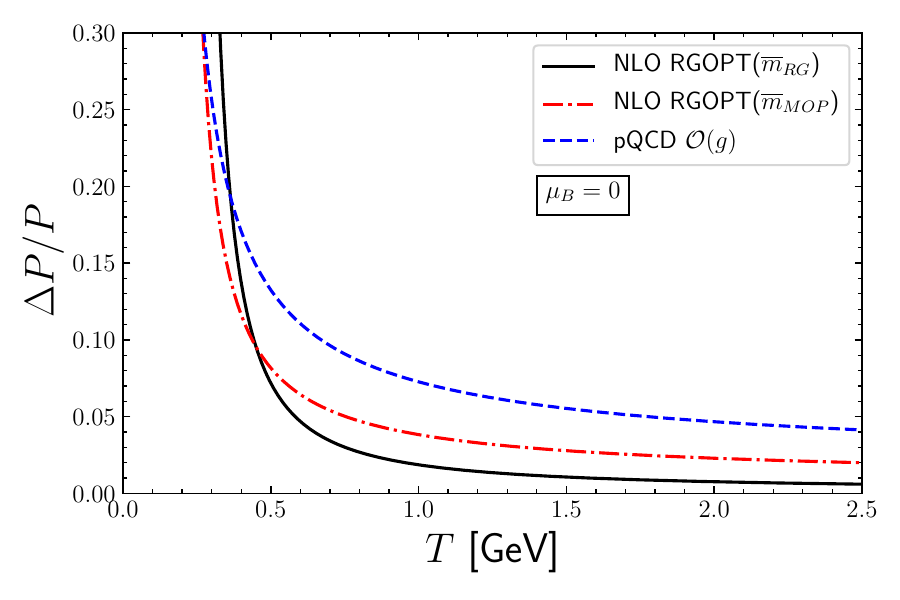,width=0.5\linewidth,angle=0}}
\caption{ $\Delta P/P \equiv P(M=4\pi T)/P(M=\pi T)-1$ as function of temperature (for $\mu_B=0$) 
for the different NLO RGOPT prescriptions compared to standard NLO pQCD, 
with scale dependence $\pi T \le M \le 4\pi T$. }
\label{DPmu0}
\end{figure}

 Similarly to Fig. \ref{Pmu0band}, we illustrate in Fig. \ref{PRGpresc} the exact NLO RGOPT pressure
as obtained from the alternative  $\overline m_{RG}$
prescription defined from solving Eqs.(\ref{RGexact}) and (\ref{B2sq0}) (explained in details in Subsec.\ref{secRG}). 
 As is seen the RGOPT reduction of remnant scale dependence is even more substantial than for the previous 
$\overline m_{MOP}$ prescription. 
The efficient reduction of remnant scale dependence with respect to standard NLO pQCD is also shown more quantitatively 
in Fig. \ref{DPmu0}, illustrating the maximal scale variations,  
$\Delta P/P \equiv (P(M=4\pi T)/P(M=\pi T)-1$, for the different approximations as indicated. 
Despite the numerically quite different MOP and RG dressed mass 
(see Fig \ref{mbarmu0}), the resulting physical pressures are much closer 
for the two prescriptions, except at very low $T$ values (i.e., very large coupling). 
This is a reasonable crosscheck of the moderate dependence upon 
the details of the optimization prescriptions, already observed here at NLO. 
For both the MOP and RG prescriptions 
lower pressure values are obtained at moderate temperatures 
as compared to LO RGOPT, NLO HTLpt and NLO pQCD in Figs \ref{Pmu0band}, \ref{PRGpresc}.
\subsection{Hot and dense quark matter}
\begin{figure}[h!]
\centerline{ \epsfig{file=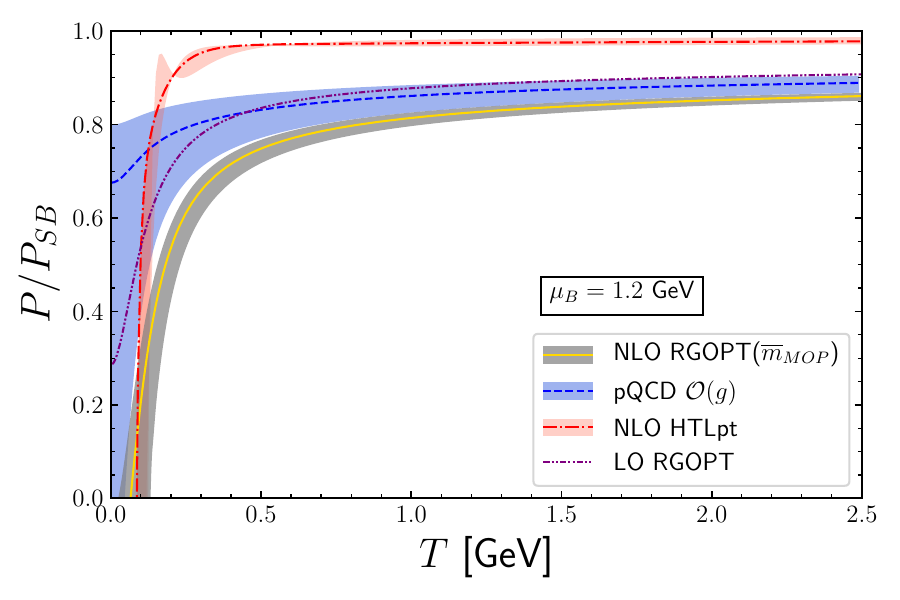,width=0.5\linewidth,angle=0}}
\caption{ RGOPT pressure as function of the temperature 
at LO and NLO (MOP prescription), compared with NLO pQCD and NLO HTLpt pressures, 
with scale variation $\pi \sqrt{T^2+ \mu^2/\pi^2} \le M \le 4\pi \sqrt{T^2+ \mu^2/\pi^2}$ at $\mu_B=1.2$ GeV. }
\label{PMOPmu400}
\end{figure}
%%%%%%%%%%%%%%%%%%%%%%%
\begin{figure}[h!]
    \centerline{ \epsfig{file=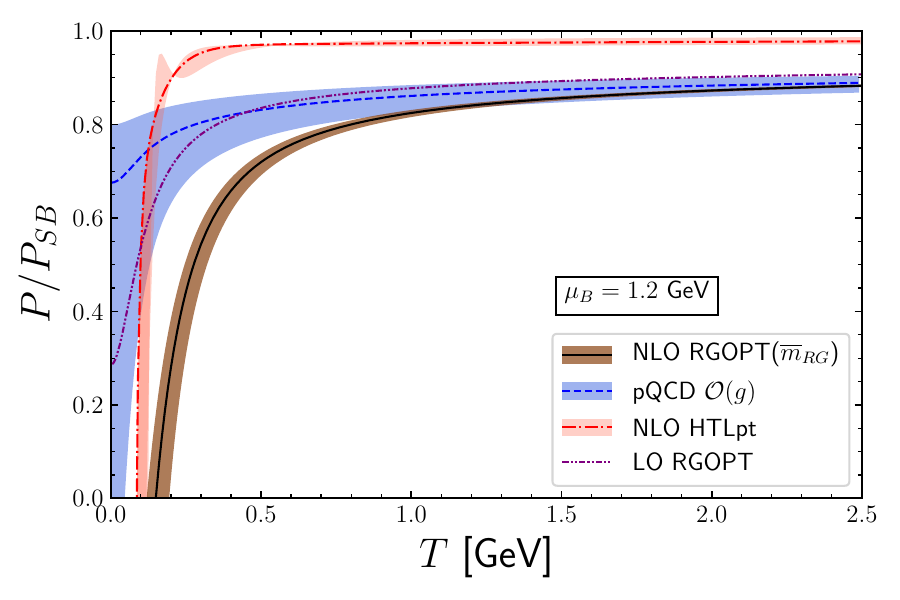,width=0.5\linewidth,angle=0}}
\caption{Same captions as in Fig. \ref{PMOPmu400} with alternative NLO RG prescription.}
\label{PRGmu400}
\end{figure}
%%%%%%%%%%%%%%%%%%%%%%%
We now consider a nonzero chemical potential values. Since the MOP (\ref{OPTexact}), (\ref{B2simp}) 
and RG  (\ref{RGexact}), (\ref{B2sq0}) prescriptions are defined
quite generically they can be readily applied to the more general 
$T,\mu \ne 0$ case. As a representative 
physical value we illustrate our results for $\mu_B=1.2$ GeV.
For the renormalization scale variation range we take as is common
$\pi \sqrt{T^2 +\mu^2/\pi^2} \le M \le 4\pi \sqrt{T^2+ \mu^2/\pi^2}$ 
within the exact NLO running coupling Eq.(\ref{g2L}). 
This gives the results for the pressure as a function of temperature
as shown in Fig. \ref{PMOPmu400} and Fig. \ref{PRGmu400} for the MOP and RG
prescriptions respectively. As is seen, for this rather sizable $\mu_B$ value the qualitative picture
is very similar to the $\mu_B=0$ case above: namely 
the remnant scale dependence reduction from RGOPT is drastic as compared to pQCD, 
and sensible departures with respect to both pQCD and HTLpt are obtained 
from resummation effects at relatively low temperatures. 
These results appear to support the robustness of the RGOPT for a more reliable exploration 
of hot and dense matter.
\subsection{Including glue contribution: confrontation to lattice results}
\begin{figure}[h!]
\centerline{ \epsfig{file=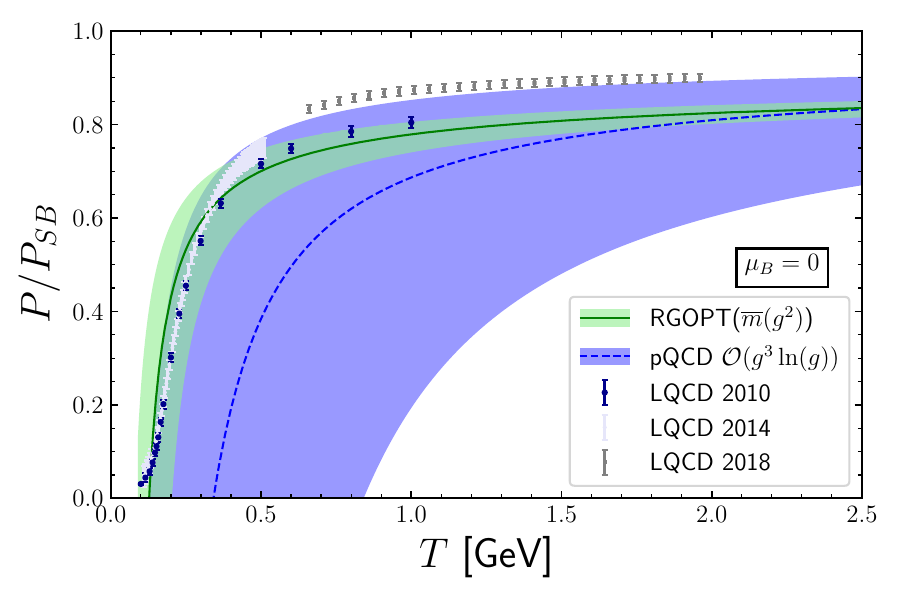,width=0.5\linewidth,angle=0}}
\caption{RGOPT $P(\overline m(g^2))$ plus NLO $P^{PT}_g$ pressure 
as function of $T$
(green band) compared to (${\rm N}^3$LO, $g^3 \ln g$) pQCD (light blue band), 
with scale dependence $\pi T \le M \le 4\pi T$, and to lattice data \cite{LQCD2010,LQCD2014,LQCD2018} at $\mu_B=0$.  }
\label{Pmg2vsPTg}
\end{figure}

\begin{figure}[h!]
\centerline{ \epsfig{file=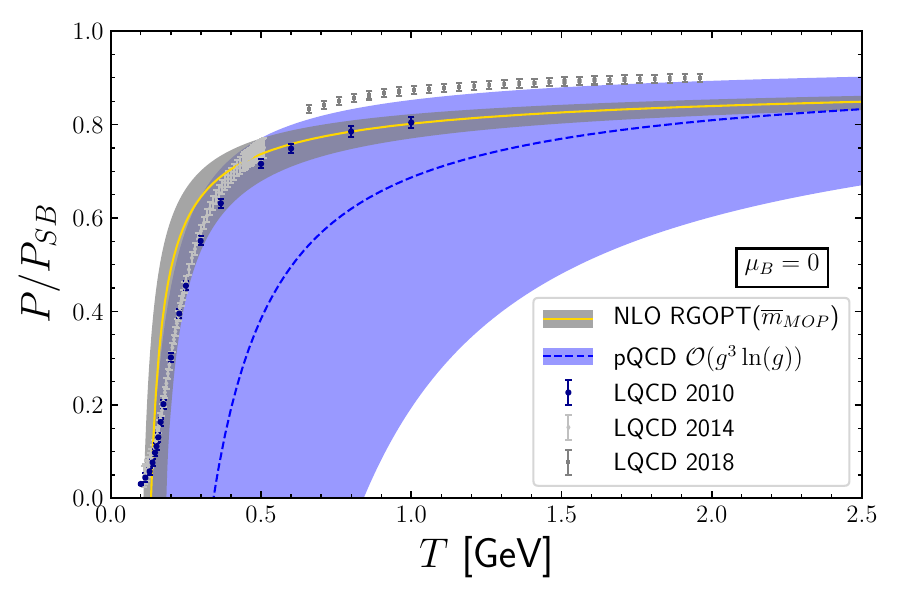,width=0.5\linewidth,angle=0}}
\caption{ Full NLO RGOPT (MOP prescription) plus NLO $P^{PT}_g$ pressure as function of $T$
(grey band) 
compared to (${\rm N}^3LO\, g^3 \ln g$) pQCD (light blue band),
with scale dependence $\pi T \le M \le 4\pi T$,  and to lattice data \cite{LQCD2010,LQCD2014,LQCD2018} at $\mu_B=0$. }
\label{Pmu0bandg}
\end{figure}

\begin{figure}[h!]
    \centerline{ \epsfig{file=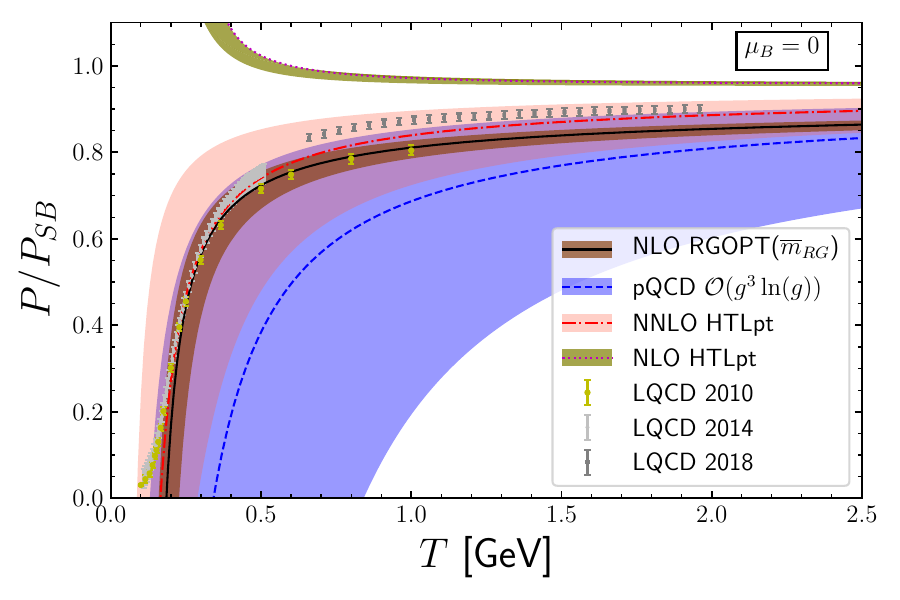,width=0.5\linewidth,angle=0}}
\caption{ Full NLO RGOPT (RG prescription) plus NLO $P^{PT}_g$ pressure (brown band) 
compared to  ${\rm N}^3{\rm LO}\, g^3 \ln g$ pQCD (light blue band),
NLO HTLpt (light green band) and NNLO HTLpt (light red band), 
with scale dependence $\pi T \le M \le 4\pi T$, and to lattice data \cite{LQCD2010,LQCD2014,LQCD2018} at $\mu_B=0$.  }
\label{PRGprescg}
\end{figure}

 In principle a rather similar RGOPT treatment 
of the pure glue sector should be possible, building on  
the hard thermal loop (HTL) originally proposed in \cite{HTLbasic},
with a gauge-invariant (non-local) effective Lagrangian  properly describing 
Landau damping and screening with a gluon (thermal) ``mass" term.  
However, this requires technically the evaluation of presently unknown and quite involved
thermal integrals.  More precisely, the RG-restoring subtraction analogous of Eq.(\ref{PRGPT}) for nonzero gluon
mass $m_D$ requires to calculate exact two-loop HTL $m^4_g \alpha_S$ contributions, rather than expanded in $m^2_D/T^2$ up to
order $\alpha_S^{5/2}$, as calculated e.g. in \cite{HTLptg2L,HTLptg3L}. Such a calculation involves
up to five-dimensional complicated integrals, due to the highly nontrivial dressing of gluon propagators and vertices rooted in the 
HTL formalism. We leave such considerations for future work~\cite{gluons}.
Therefore as above anticipated in the present work 
we treat the pure glue contribution most conservatively in a standard perturbative manner.
At the same NLO, 
the standard perturbative pure glue contribution has the well-known expression \cite{Pglue}
\be
\frac{P^{PT}_g}{P_{g,SB}} = 1 -\frac{15}{4}\left (\frac{g}{4\pi^2}\right ) +{\cal O}(g^2) ,
\label{Pg2L}
\ee
where the ideal gluon gas pressure is $P_{g,SB}=(8 \pi^2/45)\, T^4$.
Thus we simply add the perturbative NLO contribution Eq.(\ref{Pg2L}) (properly normalized) 
to our NLO RGOPT quark contributions Eq.(\ref{P2LRGOPT}), and for the numerical illustrations below we 
normalize our results to the full ideal pressure of quarks plus gluons:
$P_{SB}\to P_{q,SB}+P_{g,SB}$~\footnote{As a slight abuse of notation, note that in Figs. 
\ref{Pmg2vsPT}-\ref{PRGmu400} where
only quark contributions are included, $P_{SB}$ designates the sole quark ideal pressure Eq.(\ref{Pqsb}), while 
in Figs.\ref{Pmg2vsPTg}-\ref{PRGprescg} below $P_{SB}\equiv P_{q,SB}+P_{g,SB}$.}\\
Following the progressive elaboration levels as in the previously shown quark pressure approximations, we first
illustrate in Fig.\ref{Pmg2vsPTg} the results of using the simple perturbatively re-expanded approximation
for $\overline m$, Eq.(\ref{mbarg2}), for the quark contribution, but supplemented now by the NLO glue
contribution, Eq.(\ref{Pg2L}). The resulting RGOPT pressure is compared with both the (massless quark) 
state-of-the-art  ${\rm N}^3$LO pQCD,
which expression is taken from Ref. \cite{pQCD4L}, and to available LQCD results from 
Ref.~\cite{LQCD2010,LQCD2014,LQCD2018}.
As is seen, adding the NLO PT glue contribution puts our results
in the right ballpark of LQCD data, with clearly visible improvement as compared to pQCD, both for the central scale choice
and resulting remnant scale uncertainty. (We also note that
using instead the similar RG perturbative approximation Eq.(\ref{mbarRGg2}) gives almost undistinguishable
results from Fig. \ref{Pmg2vsPTg}, illustrating the low order perturbative consistency of the two different 
MOP and RG prescriptions).  \\
Next in Figs.\ref{Pmu0bandg} and \ref{PRGprescg}, we illustrate similarly the results obtained upon adding the NLO PT glue
contributions Eq.(\ref{Pg2L}) to the NLO RGOPT quark pressure respectively for the (exact) MOP and RG prescriptions.
These are compared with the state-of-the-art ${\rm N}^3$LO pQCD \cite{pQCD4L}, and to LQCD 
results~\cite{LQCD2010,LQCD2014,LQCD2018}.
As seen the RGOPT results get closer to LQCD data, with a further reduced scale dependence,
as compared to pQCD. 
In Fig.\ref{PRGprescg} we compare in addition with both NLO \cite{HTLptqcd2L} 
and the state-of-the-art NNLO 
HTLpt \cite{HTLptqcd3L}. The corresponding HTLpt pressure expressions are worked out from Eqs.(51),(55),(56) in \cite{HTLptqcd2L}
at NLO, and from Eqs.(4.5),(4.6) in \cite{HTLptqcd3L} at NNLO 
(we refer to Appendix C for more discussions on the HTLpt contributions). Notice also that these NNLO HTLpt and 
the ${\cal O}(g^3 \ln g)$ pQCD\cite{pQCD4L} results were obtained using a standard perturbative three-loop order running coupling.
The pressure from the RG prescription gives the smallest residual scale uncertainties, and
 is in remarkable agreement with LQCD data in \cite{LQCD2010}
for the central scale $M = 2\pi T$, 
for temperatures as low as $T \sim 0.25$ GeV
up to $T = 1$ GeV, the highest value considered in \cite{LQCD2010}. (More precisely let us mention that 
for the five available LQCD points in \cite{LQCD2010} with $T> 0.3 \,{\rm GeV}$ the central scale 
agreement is at the few permille level, and even slightly better when considering their estimated continuum data). 
It is also in good agreement  
with more recent LQCD data~\cite{LQCD2014} at intermediate $T$. 
The RGOPT pressure is somewhat closer to LQCD results from \cite{LQCD2010} than the NNLO HTLpt pressure for 
$0.5\, {\rm GeV}\lesssim T \lesssim 1 \,{\rm GeV}$, while at higher $T$ values
HTLpt is nearer to the results of \cite{LQCD2018}, and 
RGOPT shows more sizeable differences of order $5-7\%$. A concomitant feature however is the 
visible tension between low~\cite{LQCD2010} and higher $T$~\cite{LQCD2018} LQCD data in their common 
temperature range~\footnote{We show LQCD data  
as given in publicly available files\cite{LQCD2010,LQCD2014,LQCD2018}, that do not include systematic uncertainties.}. 

 Let us briefly mention that we have tried some variants of our prescriptions
in order to check the stability of our results. First,
the other RSC prescription to recover real solutions, mentioned above in Subsec.~\ref{sec2L} and used in Ref.\cite{prdCOLD},  
is to require the collinearity of the vectors
tangent to the MOP and RG curves considered as functions of $(m, g)$ (see Eq.(4.7) of Ref.\cite{prdCOLD}).
In the present $T\ne 0$ case it is however numerically much more involved 
than our simpler prescriptions above (in particular to identify the AF-compatible solutions at moderate and low $T$ values).
Yet we could check that the resulting pressure is roughly similar to the one given by the MOP prescription in Figs. \ref{Pmu0band}, \ref{Pmu0bandg}.
Next, we have also considered a variant of the RG prescription, by including the NNLO $\sim g\, m^4\, s_2$ subtraction term
of Eq.(\ref{PRGPT}), that is formally of NLO ${\cal O}(g)$~\footnote{ This variant is the 
next order analogue of including the NLO coefficient $s_1 \ne 0$ within LO RGOPT, see e.g. Eq.(\ref{mbar1L}).}. 
The $s_2$ expression \cite{JLcond,JLcond2} incorporates three-loop order
RG coefficient dependence, thus for consistency we took a three-loop perturbative
running coupling generalizing Eq.(\ref{g2L}). We remark that the resulting pressure for this variant 
hardly shows visible differences with Figs.~\ref{PRGprescg}, reflecting a good stability, 
so that we omit to illustrate it.

\begin{figure}[h!]
    \centerline{ \epsfig{file=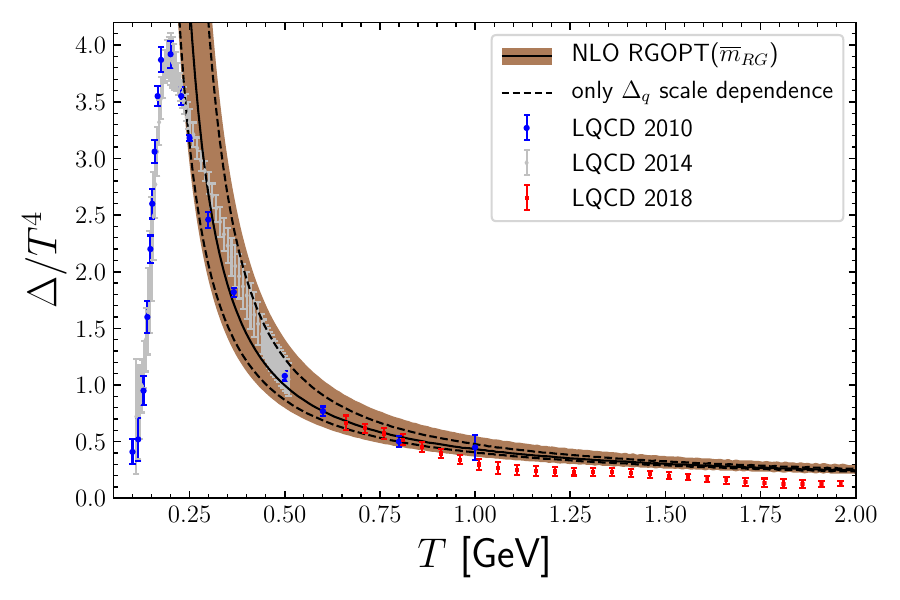,width=0.5\linewidth,angle=0}}
\caption{ NLO RGOPT (RG prescription) trace anomaly $\Delta\equiv \varepsilon -3P$ 
(including $\Delta^{PT}_g$) (brown band) 
compared to lattice data \cite{LQCD2010,LQCD2014,LQCD2018}. 
The additional dashed lines illustrate the scale uncertainty originating solely
from RGOPT quark contributions within the full scale uncertainty  added by $\Delta_g^{PT}$ (brown) band.}
\label{TraceRG}
\end{figure}

\begin{figure}[h!]
    \centerline{ \epsfig{file=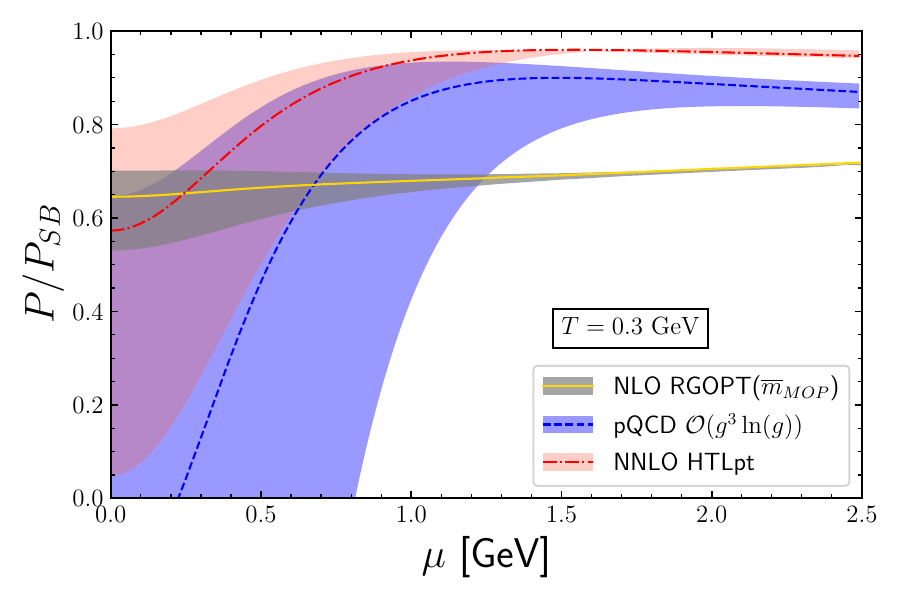,width=0.5\linewidth,angle=0}}
\caption{ Full NLO RGOPT (MOP prescription) plus NLO $P^{PT}_g$ pressure (grey band), as function of the quark chemical potential $\mu$ 
for $T=0.3$ GeV, compared to  ${\rm N}^3{\rm LO}\, g^3 \ln g$ pQCD (light blue band) and NNLO HTLpt (light red band), 
with scale dependence $\pi (T^2+\mu^2/\pi^2)^{1/2} \le M \le 4\pi (T^2+\mu^2/\pi^2)^{1/2} $.}
\label{PmuT300mop}
\end{figure}

\begin{figure}[h!]
    \centerline{ \epsfig{file=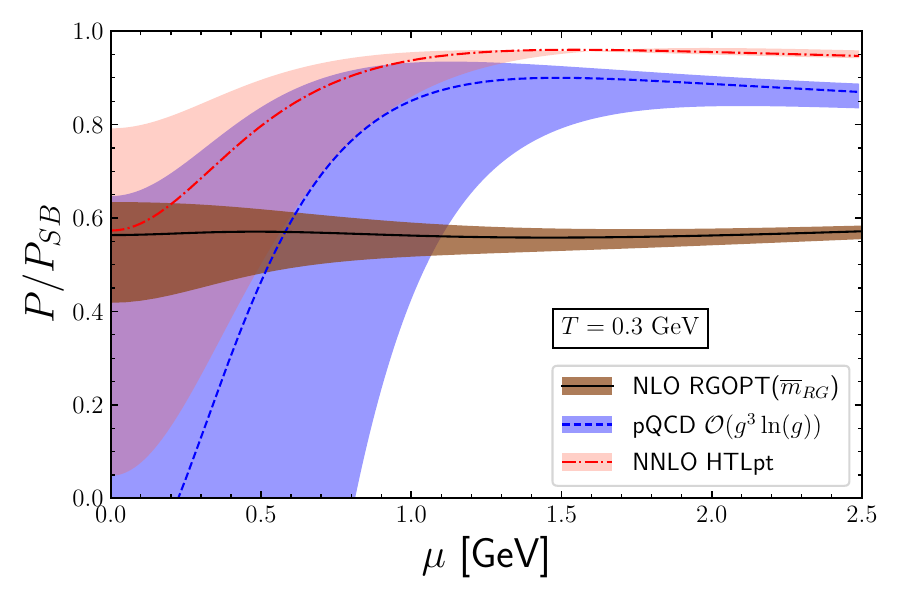,width=0.5\linewidth,angle=0}}
\caption{ Same captions as Fig. \ref{PmuT300mop} but for $\overline m_{RG}$ prescription (brown band).}
\label{PmuT300rg}
\end{figure}

Another physical quantity of interest is
the trace anomaly (or equivalently interaction measure). The latter has the well-known expression
\be
\Delta\equiv \varepsilon -3 P = T \frac{\partial P}{\partial T} -4 P = T^5\,\partial (P/T^4)/\partial T, 
\label{traceanom}
\ee
(where the second and third equalities are of course valid only for $\mu=0$). As previously
we add the pure glue NLO PT expression to our RGOPT quark contribution. The result  is illustrated,
for our best RG prescription, 
in Fig.\ref{TraceRG} where it is compared to LQCD data\cite{LQCD2010,LQCD2014,LQCD2018} only.
 A very good agreement with LQCD results of \cite{LQCD2010,LQCD2014} 
is obtained for $0.3 \,{\rm GeV} \lesssim T \lesssim 1 \,{\rm GeV} $, 
while there are more visible differences with the higher $T$ results from \cite{LQCD2018}. 
Just for indication is also delineated the part of the remnant scale uncertainties originating solely 
from the RGOPT quark contributions (dashed lines) within the total uncertainties that also include
the ones coming from the (standard)
NLO PT glue contribution.   As clearly seen however, similarly to pQCD and HTLpt, the NLO RGOPT 
does not describe correctly the peak region near the pseudocritical $T_c$ temperature as exhibited by
lattice data. We speculate that a similar RGOPT resummation in the gluon
sector should be a first necessary step to possibly better address the phase transition region, while at present 
we have treated this sector purely perturbatively as above explained. Therefore our present results are certainly
not reliable in the region  below $T \lesssim 0.25 $ GeV.  \\
As a last fairly different illustration, we show the NLO RGOPT pressure as function of the quark chemical potential $\mu$, 
for our two MOP and RG prescriptions respectively in Figs \ref{PmuT300mop}, \ref{PmuT300rg}, 
for a fixed relatively low temperature $T=0.3$ GeV, thus definitely above the ($\mu=0$) pseudocritical
temperature such that our present NLO approximation with above indicated limitations is presumably still reliable. 
As compared with HTLpt and pQCD, for both prescriptions the RGOPT pressure appears somewhat 
lower for high and moderate $\mu$ values, and exhibits a more regular behavior at low $\mu$. 
The gain in remnant scale dependence appears once more drastic. We refrain to explore 
regions closer to the transition where our present construction becomes anyway unreliable.\\ 

 To conclude this section it may be worth to recap the origin of the drastic differences between RGOPT
and HTLpt, the latter being also basically a variational modification
of the original QCD Lagrangian with mass terms, 
although based on the more elaborate HTL effective Lagrangian\cite{HTLbasic} (including among other features 
a thermal gluon mass parameter, $m_D$).
There are essentially three important differences:
\begin{itemize}
 \item 
First, the perturbative RG-restoring subtraction terms, like in Eq.(\ref{PRGPT}) typically, 
are missing in HTLpt. Accordingly the latter lacks perturbative RG-invariance formally by a 
leading order term of the massive theory pressure, ${\cal O}(m^4)\ln(M/m)$. 
Now since for any (gluon or quark) thermal masses, $m^2\sim \# g T^2$, and HTLpt is also based on high temperature expansions, 
the latter uncancelled term is effectively only a three-loop order effect,
thus largely screened and harmless at LO, and moderate even at NLO. In contrast this mismatch plainly resurfaces
at NNLO HTLpt, presumably mainly explaining the large remnant scale dependence 
observed in Refs.\cite{HTLptg3L,HTLptDense3L,HTLptqcd3L}. 
\item 
Second, the interpolating Lagrangian used in HTLpt is linear, namely with an exponent $a=1$ 
in the HTL equivalent
of Eq.(\ref{Lint}), 
instead of our RG-determined Eq.(\ref{aQCD}). As we have shown\cite{prdphi4} 
this generally spoils RG invariance even when the latter is fulfilled perturbatively by the original pressure.
\item 
Finally, remark that upon choosing a variational mass prescription Eq.(\ref{mop}) in HTLpt 
(as was done e.g. in \cite{HTLptqcd2L,HTLptDense3L}), nonreal $\overline m$ may occur,  
similarly to what happens for RGOPT (although it happens rather at NNLO in HTLpt).
In NNLO HTLpt applications this issue is avoided simply by 
replacing the gluon $\overline m_D$
arbitrary mass by a perturbative thermal mass \cite{HTLptg3L,HTLptqcd3L}, 
and taking the quark mass $\overline m_q=0$.
However, enforcing perturbative masses
is partly lacking the behavior  beyond standard perturbation potentially provided by   
more variational prescriptions.
\end{itemize}
%
%\bigskip 
%\newpage
%
\section{Conclusions and perspectives}
We have applied our RGOPT resummation approach at NLO at finite temperature and density
for the QCD quark matter.
As explained it generates more nonperturbative 
approximations with consistent RG properties already at LO (one-loop). 
Our NLO results have been compared to NLO and state-of-the-art ${\rm N}^3\mbox{LO}$ pQCD predictions
 as well as to the state-of-the-art (NNLO) HTLpt results.
Scale variations in the range $\pi T \le M \le 4\pi T$ show that at NLO the method reduces  
scale uncertainties drastically as compared to pQCD. 
Since RG properties are consistently embedded within the RGOPT, we stress that 
generically the scale uncertainty bands observed at NLO should further shrink by considering the NNLO, ${\cal O}(g^2)$.

Our two possible `MOP' or `RG' prescriptions reflect the 
often non-uniqueness of variational approaches, 
although here their respective solution is unique from 
the compelling AF-matching requirement. Moreover the visible prescription
difference for the resulting dressed mass (see Fig.\ref{mbarmu0}) is perturbatively consistent at low orders 
(Eqs.(\ref{mbarg2}), (\ref{mbarRGg2})), and 
is substantially reduced within the resulting physical pressures.  
Using the RG Eq.(\ref{RGred})
prescription, that more directly embeds consistent RG properties, not surprisingly 
gives the best remnant scale dependence at NLO (at it also happened in other considered models \cite{prdphi4}).
Once a specific RSC is adjusted
to recover real solutions, 
the discrepancies between possibly different RSC 
prescriptions are formally perturbatively higher order terms. 
Nevertheless since we 
consider all expressions exactly rather than perturbatively truncated, 
numerically the RSC has a moderate net effect on the final pressure results.
As we have illustrated, 
any perturbative reexpansion of the exact solutions somehow degrades the scale dependence.

Concerning the full QCD pressure, due
to present technical limitations in applying the RGOPT plainly to
the glue sector, in this work we have adopted a simple-minded 
approach, adding the purely perturbative NLO glue contributions
to the pure quark sector resummed by RGOPT. 
We have confronted the resulting
predictions for the QCD pressure with available LQCD results. For our best RG prescription 
the central scale $M = 2\pi T$ results are in remarkable agreement with the LQCD 
results \cite{LQCD2010,LQCD2014} 
for temperatures as low as $T \gtrsim 0.25$ GeV, which lies
within the nonperturbative regime, up to $T = 1$ GeV. 
 However, similarly to pQCD and HTLpt, the NLO RGOPT construction 
explored in the present work is unable to describe the peak region near the pseudocritical $T_c$ temperature as exhibited by
lattice data. Although our simple prescription appears to describe fairly well the moderate to high-$T$ regimes 
$T \gtrsim 0.25 \, {\rm GeV} \sim 1.5\, T_{pc}$, going beyond NLO one would not avoid
to face the infrared divergences from gluon contributions, calling for appropriate resummations.
The striking matching with LQCD results from Ref. \cite{LQCD2010} 
as seen in Fig.\ref{PRGprescg} may be partly numerically accidental, but variants of our
prescription, specifically the MOP pressure in Fig.\ref{Pmu0bandg}, still appears 
in very good agreement given our essentially NLO construction.
Moreover the RG properties native to the RGOPT are not accidental in 
drastically reducing the scale dependence problem, particularly 
when comparing our NLO results to NNLO HTLpt. 
 There are however some visible differences between our results and  
higher $1 \, {\rm GeV}\lesssim T \lesssim 2\, {\rm GeV}$ LQCD data\cite{LQCD2018}.
We remark that the LQCD pressure results 
in \cite{LQCD2010} and in Ref. \cite{LQCD2018}
appear to be in tension in their common temperature range, while the trace anomaly 
shows more continuity~\footnote{ The LQCD simulations in Refs. \cite{LQCD2010,LQCD2014,LQCD2018} 
primarily calculate the trace anomaly $\Delta$, the pressure being derived by the integral method, 
i.e. essentially from numerically integrating the last equality 
in Eq.(\ref{traceanom}).}, a feature that may call for more investigations independently of our results.
When comparing with $2+1$ flavor LQCD as here illustrated, 
one may also keep in mind our presently not fully realistic
approximation of $N_f= 3$ degenerate flavors.
 As illustrated the RGOPT properties extend without much degradation to sizable chemical potential values and 
relatively low temperatures (see Figs. \ref{PMOPmu400}, \ref{PRGmu400}, and Figs. \ref{PmuT300mop}, \ref{PmuT300rg}), 
that indicates the potential of our approach towards a more systematic exploration of hot and dense matter.
Future applications may consider the inclusion of physical quark masses to generate a more realistic 
 equation of state.
\acknowledgments 
 We thank Peter Petreczky for bringing the results of Ref.~\cite{LQCD2018} to our attention.
 We thank Eduardo Fraga and Rudnei Ramos for related discussions.
M.B.P. is partially supported by Conselho Nacional de Desenvolvimento Cient\'{\i}fico e Tecnol\'{o}gico (CNPq-Brazil), 
Process No. 303846/2017-8, and by Coordena\c c\~{a}o  de Aperfei\c coamento de Pessoal  de  N\'{\i}vel 
Superior-(CAPES-Brazil)-Finance  Code 001. This author also thanks the Charles Coulomb Laboratory, in Montpellier, 
for the hospitality. T.E.R. thanks the support and hospitality of CFisUC where part of this work was developed 
and acknowledges Conselho Nacional de Desenvolvimento Cient\'{\i}fico e Tecnol\'{o}gico (CNPq-Brazil) and 
Coordena\c c\~{a}o  de Aperfei\c coamento  de  Pessoal  de  N\'{\i}vel  Superior   (CAPES-Brazil) for 
PhD grants at different periods of time. This  work  was  financed  in  part  by   
INCT-FNA (Process No.  464898/2014-5).

\bigskip

\appendix 

\section{high-$T$ limit}\label{ApphT}
We give here for completeness the well-known $T\gg m, \mu=0$ approximations (see e.g.\cite{laine,kapusta-gale})
of the basic thermal integrals defined in Eqs~(\ref{J1def})-(\ref{J3def}):
\begin{equation}
2  J_1(T\gg m,\mu=0) \approx \frac{7 \pi^2}{180} -\frac{m^2}{12\,T^2}  + \frac{m^4}{T^4} \frac{2}{(4\pi)^2} 
\left [  \frac{3}{4} -\ln \left( \frac{m e^{\gamma_E}}{\pi T} \right )  \right ] +{\cal O}\left (\frac{m^6}{T^6} \right)\;,
\end{equation}
\begin{equation}
J_2 (T\gg m,\mu= 0) \approx  \frac{1}{12} +\frac{1}{4\pi^2} \frac{m^2}{T^2} 
\left [ \ln \left ( \frac{m e^{\gamma_E}}{\pi T} \right ) - \frac{1}{2} \right ] +{\cal O}\left (\frac{m^4}{T^4} \right )\;,
\label{J2hT}
\end{equation}
and the more complicated genuine two-loop integral $J_3$ of Eq.(\ref{J3def}) has a finite $m\to 0$ limit (however
not analytically integrable to our knowledge, we give below its numerically integrated approximate value):
\begin{equation}
J_3\left (\frac{m}{T}\to 0,\frac{\mu}{T}=0 \right ) = \frac{4}{(2\pi)^4}\int_0^\infty d \hat p \int_0^\infty d \hat q \,n_F(\hat p) n_F(\hat q)\,
\ln \left(\frac{|\hat p-\hat q|}{\hat p+\hat q}\right) 
 +{\cal O}\left (\frac{m^2}{T^2} \right )\simeq
-0.00129532 +{\cal O}\left (\frac{m^2}{T^2} \right )
\end{equation}
where $\hat p, \hat q\equiv p/T, q/T$ and $n_F(p)= (e^p+1)^{-1}$ is the Fermi-Dirac distribution.
\section{Numerical $\overline m $ solutions at NLO}\label{Appnumsol}
We discuss here in some details the behavior of the exact NLO numerical solutions for the two MOP or RG prescriptions
as defined in Secs.\ref{secMOP}, and \ref{secRG}.
Note that using directly the MOP Eq.(\ref{OPTexact}) or the RG Eq.(\ref{RGexact}) 
makes the AF solution identification obvious.
Concerning the MOP Eq.(\ref{OPTexact}), once $B_2$ is consistently determined by Eq.(\ref{B2simp}) such as to recover 
$D_{mop}>0 $ in Eq.(\ref{Delta}),
one sees from the structure of (\ref{OPTexact}) that $(-)$ (AF) solutions only exist if 
$-\ln (m^2/M^2) +B_{mop} >0$, and conversely $(+)$ (non-AF) solutions only exist if 
$-\ln (m^2/M^2) +B_{mop} <0$.
%%%%%%%%%%
\begin{figure}[h!]
    \centerline{ \epsfig{file=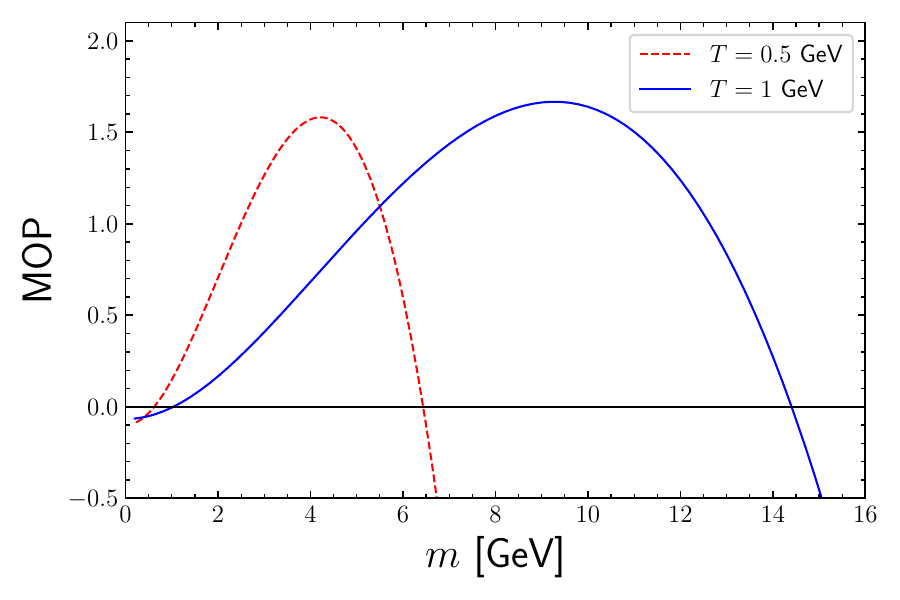,width=0.5\linewidth,angle=0}}
\caption{ AF and non-AF roots of MOP Eqs.(\ref{OPTexact}), (\ref{B2simp}) for $M=2\pi T$ and 
for two representative $T$ values, $T=0.5$ GeV (dashed), $T=1$ GeV (thick).}
\label{OPTroot}
\end{figure}
%%%%%%%%%%%%%
Once $M, g(M)$ are taken to be $T,\mu$-dependent via the perturbative running coupling Eq.(\ref{g2L}),   
Eq.(\ref{OPTexact}) becomes a function of $m/T$ and $g(T/\Lambda_{\overline{MS}})$. 
Despite the nonlinear dependence in $m/T$, at the level of Eq.(\ref{OPTexact}) both the AF and non-AF solutions happen 
to be unique in their respective existence range. This is illustrated in Fig. \ref{OPTroot} (for $\mu=0$)
for two representative low to moderate temperatures, respectively $T=0.5$ and $T=1$ GeV, and for the central scale choice $M=2\pi T$. 
It is also clear that for any $T$ the smallest solution is the AF one:
Indeed for $g(\pi T \le M\le 4\pi T)$, $-\ln (m^2/M^2) +B_{mop} $ is a monotonically decreasing function of 
$m$ for fixed $T$, 
and is $>0$ (respectively $<0$) below (respectively above) a given $m_0$, such 
that necessarily $\overline m(\mbox{AF}) < m_0 < \overline m(\mbox{non-AF})$. 
The value of $m_0$ depends quite strongly on $T$ (and $M$):
typically for the input corresponding to Fig. \ref{OPTroot} with $M=2\pi T$, one finds $m_0\simeq 1.28 (1.91)$ for  
$T=0.5 (1.0)$ GeV respectively. 
(Notice also that in Fig. \ref{OPTroot} the non-AF solution is 
unrealistically large with respect to $T$, that also makes it easy to unambiguously select 
the correct AF-matching solutions).

At $\mu=0$,
following the AF-matching $\overline m$ of Eq.(\ref{OPTexact}) continuously from $T=0$ to arbitrary $T$ is in 
principle possible,
although only for a fixed scale $M$ (thus a fixed $g(M)$) unrelated to $T$, otherwise obviously
at some small $M\sim \pi T$ one hits on $M\sim \Lambda_{\overline{MS}}$ where the perturbative coupling
diverges. For sizable $\mu\ne 0$ the latter problem if avoided if defining as conventional $M\sim \pi \sqrt{T^2+\mu^2/\pi^2}$
(provided that one is not in the case of both $T \ll \mu $ and small $\mu$).

Finally concerning the RG Eq.(\ref{RGexact}), both NLO solutions are already AF-matching, giving thus a 
unique solution upon using the 
prescription Eq.(\ref{B2sq0}). Numerically the exact $\overline m_{RG}$ 
solution of Eq.(\ref{RGexact}) is somewhat larger than $\overline m_{MOP}$ for a given $T$, as 
illustrated in Fig. \ref{mbarmu0}.
\section{NLO and NNLO HTLpt expressions}\label{AppHTL}
For completeness we specify here how 
the NLO\cite{HTLptqcd2L} and NNLO\cite{HTLptqcd3L} HTLpt pressure expressions
were precisely used when compared with other results. In particular for consistent comparison purposes in 
Figs.~\ref{Pmg2vsPT},\ref{Pmu0band}, and Figs.~\ref{PMOPmu400},\ref{PRGmu400} 
we aim to pin down
the HTLpt equivalent of the sole quark contributions, as shown up to NLO in Fig. \ref{Fig0}, but with the quark and gluon
propagators and quark-gluon vertex replaced with HTL-dressed ones consistently.
More precisely from first comparing Eq.(51) of \cite{HTLptqcd2L} to the pure glue NLO HTLpt pressure 
(given e.g. in Eq.(4.8) of second Ref. in \cite{HTLptg3L}), it is not difficult to single out all terms
originating solely from the pure quark vacuum energy. Next, from the resulting pressure we have rederived
the (variationally determined) dressed thermal gluon $m_D$ and quark $m_q$ mass as in Eqs.(55),(56) in \cite{HTLptqcd2L},
that amounts to remove in these expressions the pure glue contributions (terms $\propto c_A$ in Eq.(55)
in \cite{HTLptqcd2L}).\\ 
At NNLO of HTLpt, a well-defined separation between pure quark and pure glue
contributions appear ambiguous as these become more entangled. 
When comparing with our complete QCD RGOPT pressure
e.g. in Fig.\ref{PRGprescg} and subsequent figures, we obviously took the complete QCD NNLO HTLpt pressure,
as given e.g. in Eqs.(4.5),(4.6) of Ref.\cite{HTLptqcd3L} (see also Ref.\cite{HTLptDense3L}).

\end{document}